\pdfoutput=1
 
\documentclass[conference,fleqn]{IEEEtran}
\usepackage[cmex10]{amsmath}
\usepackage{amssymb}
\usepackage{amsthm}
\usepackage{mathtools}
\usepackage[pdftex]{graphicx}
\usepackage[font=footnotesize]{subfig}
\usepackage{cite}
\usepackage{flushend}
\usepackage{booktabs}

\usepackage{hyperref}
\hypersetup{%
  pdftitle={Maximum Entropy Models of Shortest Path and Outbreak Distributions in Networks},
  pdfauthor={Christian Bauckhage, Kristian Kersting, Fabian Hadiji},
  pdfsubject={network science, network analysis, graph theory},
  pdfkeywords={shortest path distributions, epidemic spreading, generalized Gamma distribution},
  colorlinks=true,
  urlcolor=blue,
  citecolor=green,
  bookmarks=false}

\renewcommand{\vec}{\boldsymbol}
\newcommand{\mat}[1]{\boldsymbol{#1}}

\newcommand{\gashp}{\eta}
\newcommand{\wbscl}{\lambda}
\newcommand{\wbshp}{\kappa}
\newcommand{\ggscl}{\sigma}
\newcommand{\ggshpa}{\alpha}
\newcommand{\ggshpb}{\beta}

\graphicspath{{./}}

\begin{document}

\title{Maximum Entropy Models of Shortest Path \\ and Outbreak Distributions in Networks}

\author{%
\IEEEauthorblockN{Christian Bauckhage}
\IEEEauthorblockA{Fraunhofer IAIS \\ 53754 St.Augustin, Germany}
\and
\IEEEauthorblockN{Kristian Kersting}
\IEEEauthorblockA{TU Dortmund University \\ 44227 Dortmund, Germany}
\and
\IEEEauthorblockN{Fabian Hadiji}
\IEEEauthorblockA{TU Dortmund University \\ 44227 Dortmund, Germany}
}

\maketitle

\begin{abstract}
Properties of networks are often characterized in terms of features such as node degree distributions, average path lengths, diameters, or clustering coefficients. Here, we study shortest path length distributions. On the one hand, average as well as maximum distances can be determined therefrom; on the other hand, they are closely related to the dynamics of network spreading processes. Because of the combinatorial nature of networks, we apply maximum entropy arguments to derive a general, physically plausible model. In particular, we establish the generalized Gamma distribution as a continuous characterization of shortest path length histograms of networks or arbitrary topology. Experimental evaluations corroborate our theoretical results.
\end{abstract}

\section{Introduction}

As networks are combinatorial structures, network analysis typically relies on statistics such as node degree distributions, average path lengths, clustering coefficients, or measures of assortativity \cite{Cohen2010-CN}. Histograms of shortest path lengths are considered less frequently, but they, too, provide useful information. First of all, features such as average path lengths or network diameters can be determined therefrom. Second of all, path length statistics are closely related to velocities or durations of network spreading processes. 
Analytically tractable models of shortest path distributions would thus allow for inference of network properties as well as for reasoning about diffusion dynamics.

Yet, although shortest path length distributions are an easy enough concept, analytic models prove difficult to obtain and related work is scarce \cite{Bauckhage2013-TWA,Blondel2007-DDI,Fronczak2004-APL,Ukkonen2014-IEO,Vazquez2006-PGI}. Here, we contribute to these efforts and derive a new general model of shortest path length distributions in strongly connected networks. Considering a duality between network spreading processes and shortest path lengths, we show that maximum entropy arguments as to the dynamics of diffusion processes lead to the generalized Gamma distribution.

\begin{figure}[t!]
\centering
\subfloat[$t=0$]{\includegraphics[width=0.48\columnwidth]{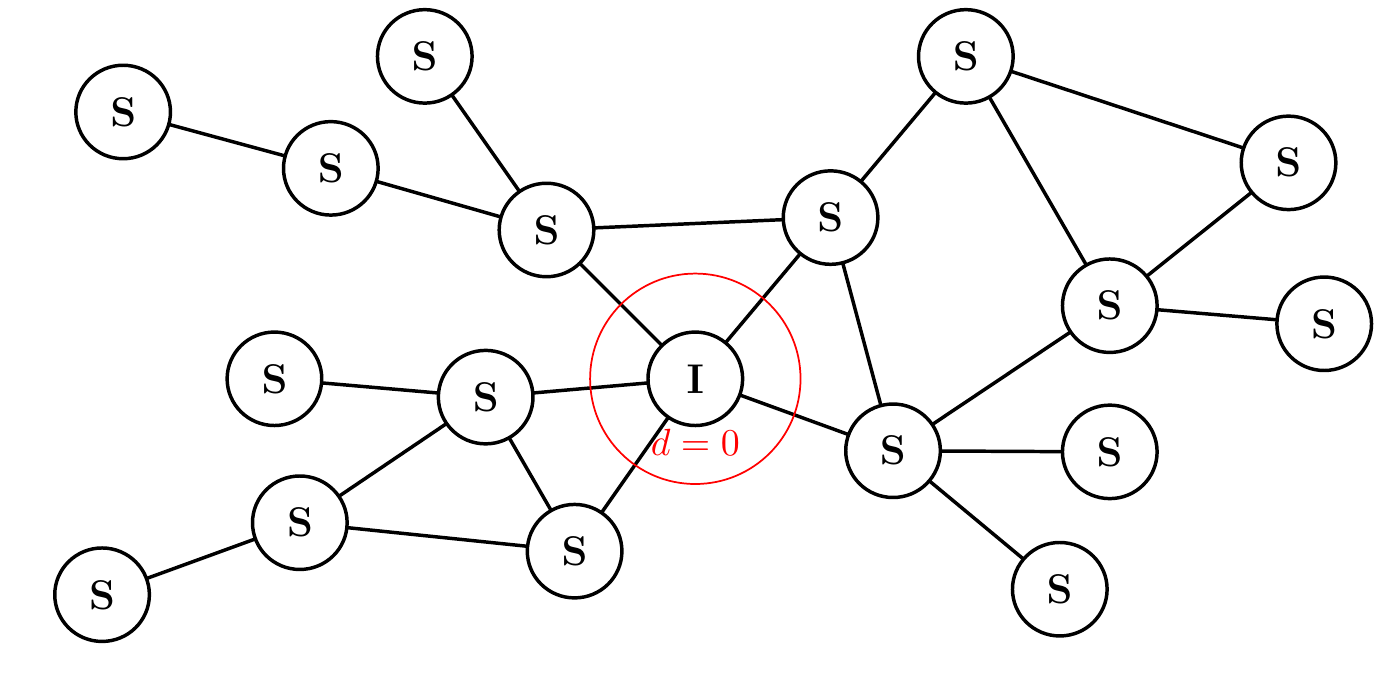}}{\ }
\subfloat[$t=1$]{\includegraphics[width=0.48\columnwidth]{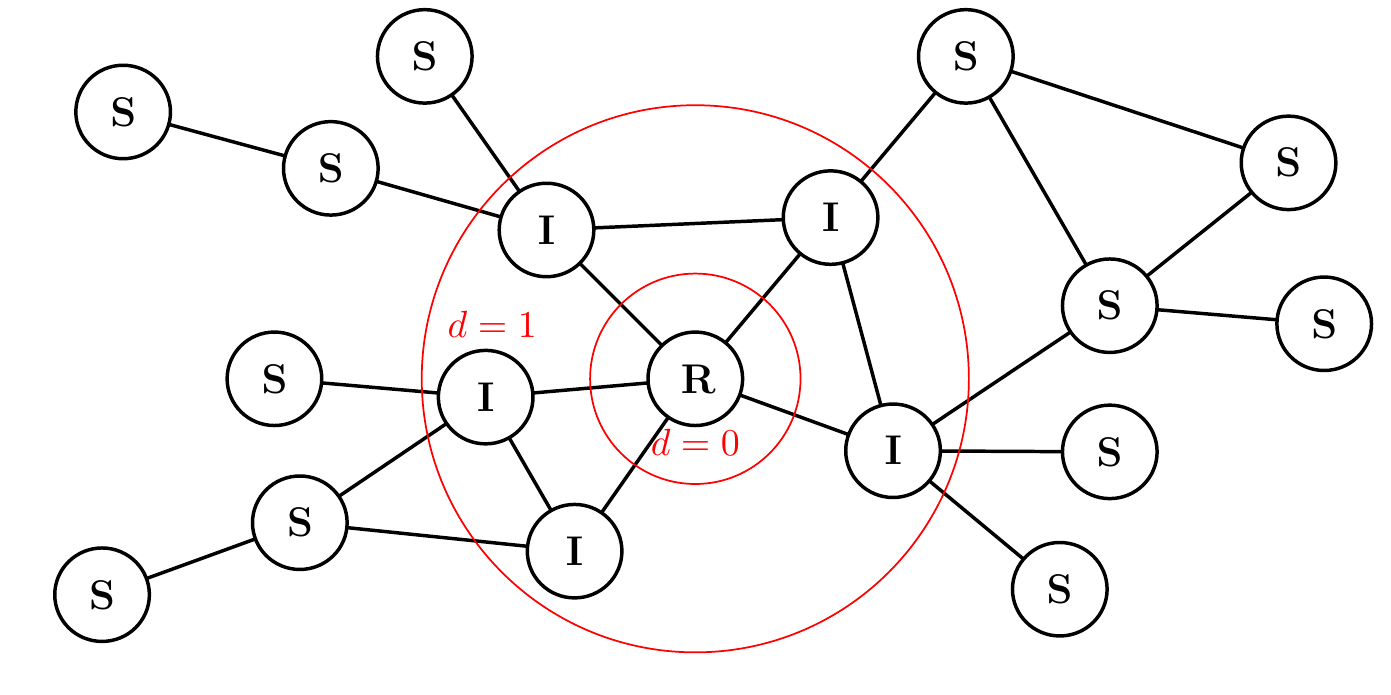}}

\subfloat[$t=2$]{\includegraphics[width=0.48\columnwidth]{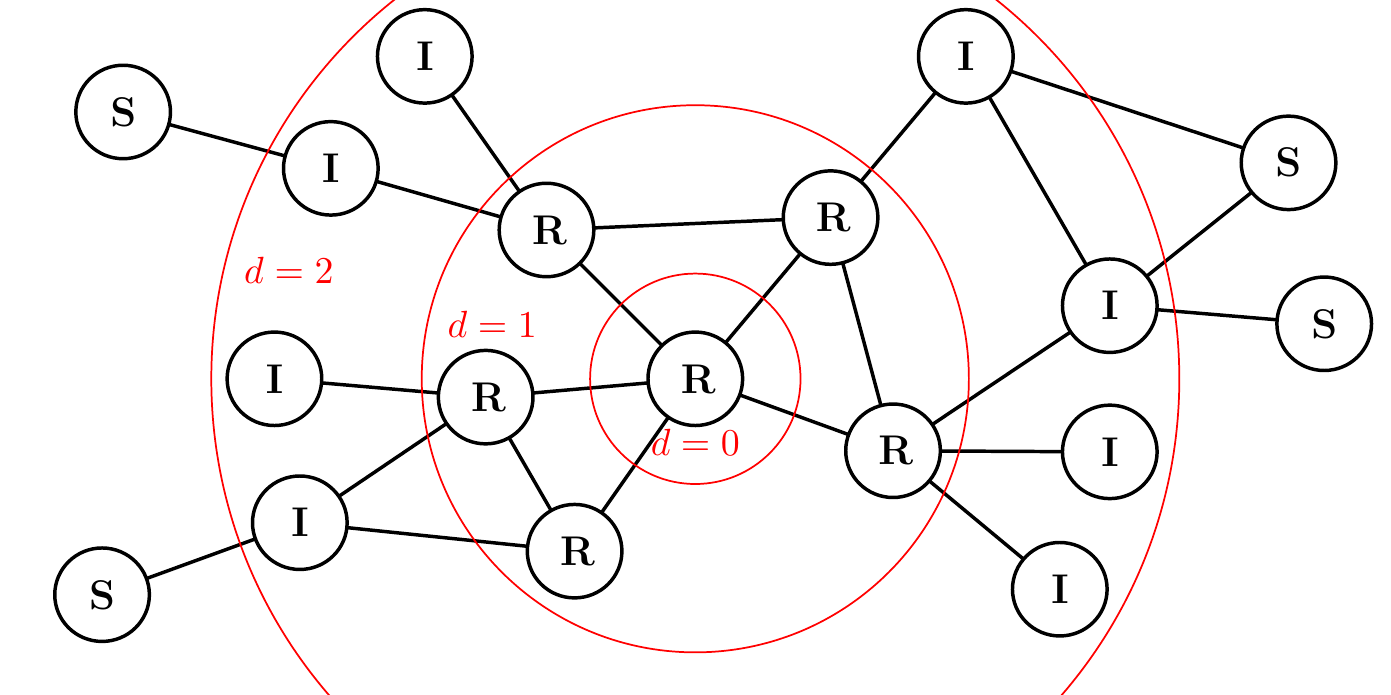}}{\ }
\subfloat[$t=3$]{\includegraphics[width=0.48\columnwidth]{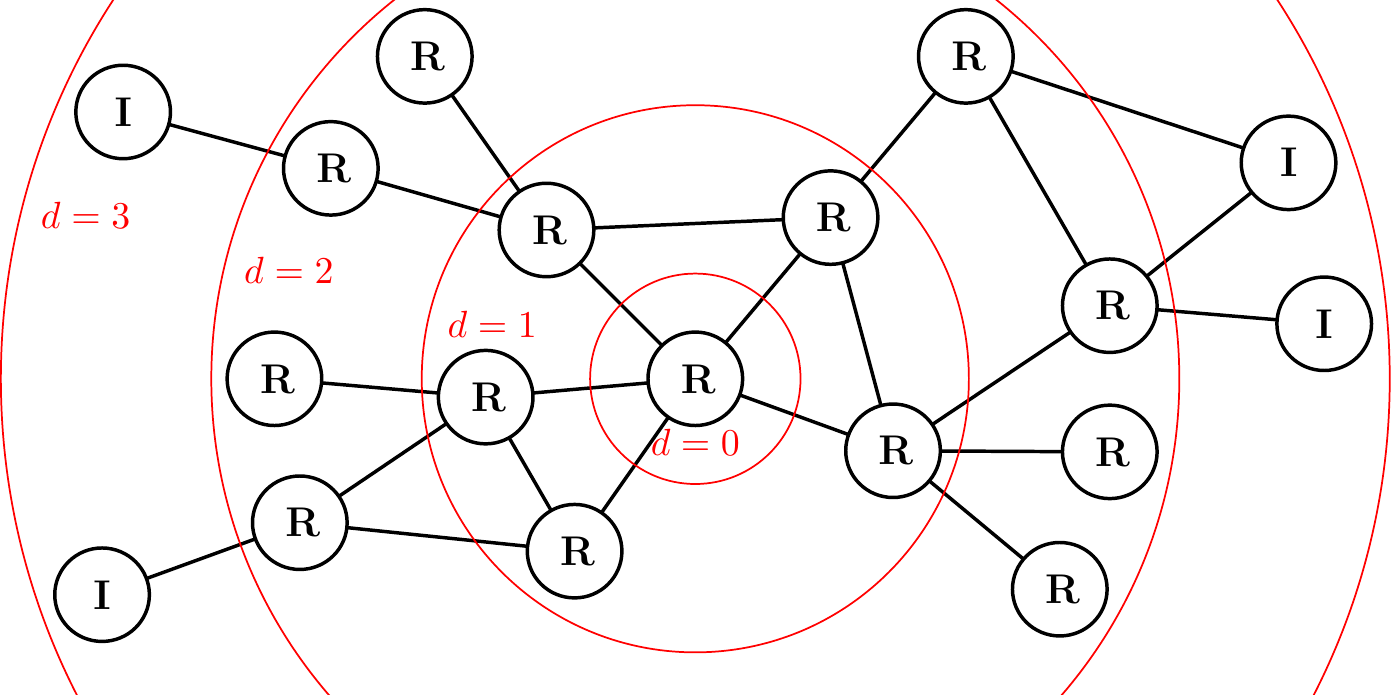}}
\caption{\label{fig:didactic-spread} A highly infectious \emph{SIR} cascade on a network realizes a node discovery process. (a) at onset time $t=0$, most nodes are \emph{susceptible} and only a single node is \emph{infected}. (b)--(d) infected nodes immediately \emph{recover} but always infect their susceptible neighbors. This way, the number of newly infected nodes $n_t$ at time $t$ corresponds to the number $n_d$ of nodes that are at distance $d=t$ from the source node of the epidemic.}
\end{figure}

\subsection{Motivation}

Models of network spreading processes play an important role in various disciplines. They characterize the dynamics of epidemic diseases \cite{Keeling2005-NAE,Lloyd2001-HVS}, explain the diffusion of innovations or viral messages \cite{Acemoglu2011-DOI,Leskovec2007-TDO}, and, in the wake of social media, a quickly growing body of research develops graph diffusion models to study patterns of information dissemination in Web-based social networks \cite{Adar2005-TIE,Bauckhage2011-III,Budak2010-LTS,Iribarren2009-IOH,Leskovec2009-MTA,Yang2011-POT}.

Each of these examples concerns an instance of a rather general phenomenon: An agent (a virus, a rumor, an urge to buy a product, etc.) spreads in form of a contact process and thus cascades through a network of interlinked entities (people, computers, blogs, etc.). At the onset of the agent's activity, many networked entities are \emph{susceptible} to its effects but only a few are actually \emph{infected} (see Fig.~\ref{fig:didactic-spread}(a)). As time progresses, susceptible entities related to infected ones may become infected whereas infected entities may remain infected, \emph{recover}, become susceptible again, or even be removed from the population (see Fig.~\ref{fig:didactic-spread}(b)--(d)).

\begin{figure}[t!]
\centering
\includegraphics[width=0.9\columnwidth]{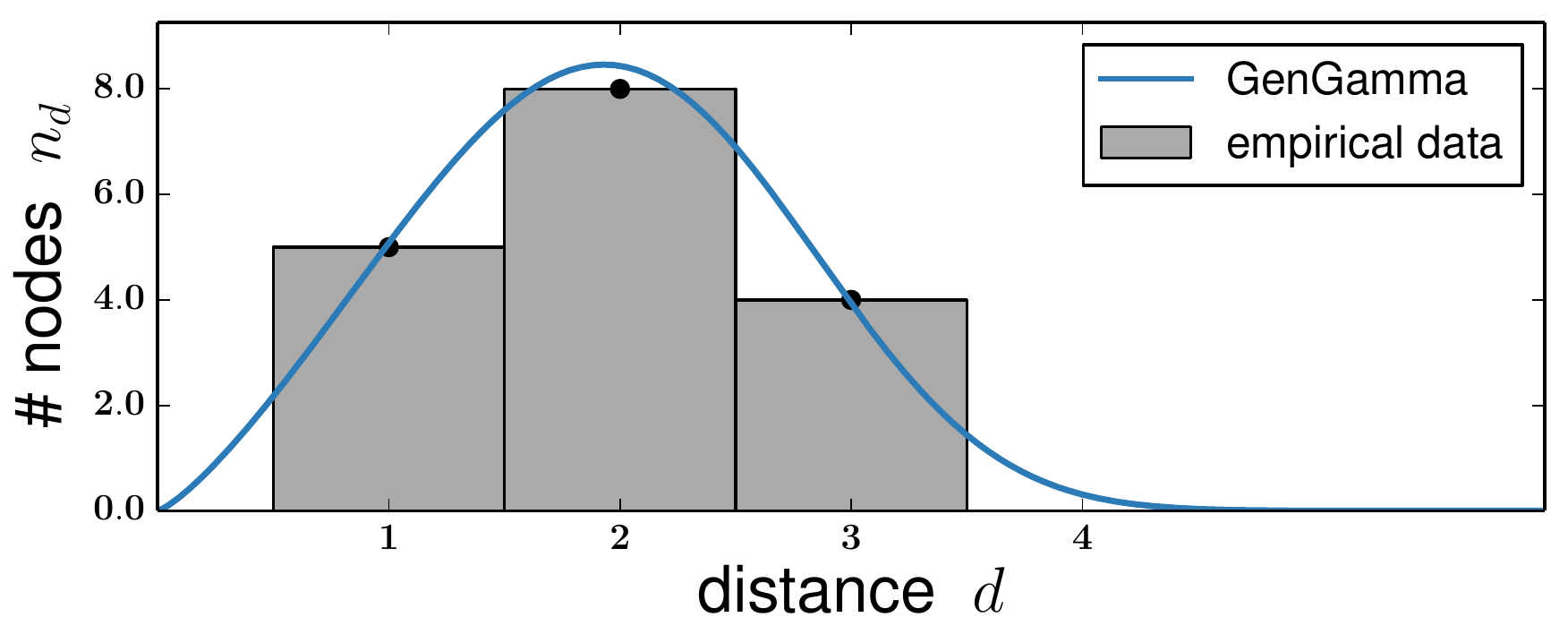}
\caption{\label{fig:didactic-hist} Shortest path histogram resulting from the network spreading process in Fig.~\ref{fig:didactic-spread} and a corresponding maximum likelihood fit of the generalized Gamma distribution.}
\end{figure}

Crucial properties of such a process are its infection rate, its recovery rate, or the number of infected entities per unit of time. If an (information) epidemic is observed to rage through a community, these features help assessing its progression or final outbreak size and thus inform contagion or dissemination strategies. For instance, while public health authorities need to devise immunization protocols to curtail epidemic diseases, viral marketers typically aim at maximizing the effects of their campaigns. In any case, knowledge as to the structure of a community through which an epidemic spreads would be beneficial. Alas, in practice, community structures are hardly ever known but available information only consists of outbreak data as shown in Fig.~\ref{fig:didactic-hist}. 

In this paper, we address the problem of relating outbreak data to network structures. Our approach is orthogonal to related contributions in \cite{Adar2005-TIE,Kempe2003-MTS,Leskovec2007-POC,Rodriguez2013-SAD,Yang2011-POT} which assume outbreak data \emph{and} information about network members to be available and apply machine learning to identify hubs, link structures, or infection routes. Instead, we follow the paradigm in \cite{Barthelemy2004-VAH,Bauckhage2013-TWA,Iribarren2009-IOH,Newman2002-TSO,Pastor2001-ESI,Vazquez2006-PGI} which asks for physical explanations of the noticeably skewed appearance of outbreak histograms and attempts to relate corresponding physical models to network properties.

\subsection{Main Results and Overview}

We establish the generalized Gamma distribution as a continuous, analytically tractable model of discrete shortest path histograms of networks of arbitrary topology. Considering the fact that temporal distributions of infection counts in highly infectious spreading processes and distributions of shortest path lengths are dual phenomena allows us to invoke maximum entropy arguments from which we derive physical characterizations of path lengths- and outbreak statistics. 

In other words, our model results from first principles rather than from data mining. It also generalizes previous theoretical results \cite{Bauckhage2013-TWA,Vazquez2006-PGI} and provides an explanation for a recent empirical observation as to path length distributions in social networks \cite{Bild2014-ACO}. 

In section~\ref{sec:model}, we briefly review existing analytical models of shortest path- and outbreak distributions, show that the generalized Gamma distribution naturally generalizes these, and discuss its properties and characteristics. 

Note that, for the sake of readability, we defer technical details on how the generalized Gamma emerges as a maximum entropy model of path length- and outbreak data to appendices \ref{sec:derivation}, \ref{sec:timescales}, and \ref{sec:alpha}.

In section~\ref{sec:experiments}, we empirically test our theoretical predictions and compare the predictive power of our new model to those of related previous models. In extensive experiments with synthesized networks of different topologies as well as with the real world networks (social, bipartite, and natural) contained in the KONECT collection \cite{Kunegis2013-KON}, we find that the generalized Gamma distribution indeed provides highly accurate fits to shortest path histograms. In addition, we simulate a large number of epidemic spreading processes on different synthetic networks and find the generalized Gamma distribution to account well for the resulting outbreak data. Finally, we present first empirical evidence that different networks topologies lead to considerably different parameters of fitted generalized Gamma distributions which hints at new approaches towards network inference.

Finally, in section~\ref{sec:conclusion}, we summarize our contributions and findings and discuss how they may inform future work on network analysis and network spreading processes.

\section{Theoretical Model}
\label{sec:model}

Analytically tractable models of shortest path length distributions and outbreak data are of considerable interest in the study of epidemic processes on networks. However, as networks are combinatorial structures, corresponding results are necessarily statistical \cite{Barthelemy2004-VAH,Bauckhage2013-TWA,Iribarren2009-IOH,Newman2002-TSO,Pastor2001-ESI,Vazquez2006-PGI}. In this section, we briefly review existing models and discuss the basic ideas and properties of our new model. In doing so, we ignore technical details but focus on overall results so that this paper is more easily accessible to a wider audience. The mathematical derivation of our main result as well as its characteristics are presented in the appendix.

\subsection{Known Results and Observations}

In a landmark paper \cite{Vazquez2006-PGI}, Vazquez studied the dynamics of epidemic processes in power law networks. Arguing based on branching process models, he showed that for networks whose node degree distribution has a power law exponent $2 < \gamma < 3$, the number of infected nodes at time $t$ follows a \emph{Gamma distribution}.
\begin{equation}
\label{eq:GA}
f_{GA}(t \mid \theta, \gashp) = \frac{1}{\theta^\gashp} \frac{1}{\Gamma(\gashp)} t^{\gashp-1} e^{-t/\theta}
\end{equation}
where $\Gamma(\cdot)$ is the gamma function and $\theta > 0$ and $\gashp > 0$ are scale and shape parameters, respectively. 
Curiously, for power law exponents $\gamma \geq 3$, this result does not apply. 

Concerned with Erd\H{o}s-R\'enyi graphs, the work in \cite{Bauckhage2013-TWA} derived a different result. Based on models of the expected number of paths of length $t$ between arbitrary nodes \cite{Blondel2007-DDI,Fronczak2004-APL}, it considered extreme value theory \cite{deHaan2006-EVT} to show that infection counts can be characterized by the \emph{Weibull distribution}.
\begin{equation}
\label{eq:WB}
f_{WB}(t \mid \wbscl, \wbshp) = \frac{\wbshp}{\wbscl^\wbshp} t^{\wbshp-1} e^{-(t/\wbscl)^\wbshp}
\end{equation}
where $\wbscl > 0$ and $\wbshp > 0$ are scale and shape parameters.

Note that neither model was obtained from mere empirical observations. Both follow from basic principles, provide \emph{physically plausible} and comprehensible characterizations of shortest path length distributions and diffusion dynamics, and were verified empirically.

\begin{figure}[t!]
\centering
\subfloat[$p = 0.005$]{\includegraphics[width=0.48\columnwidth]{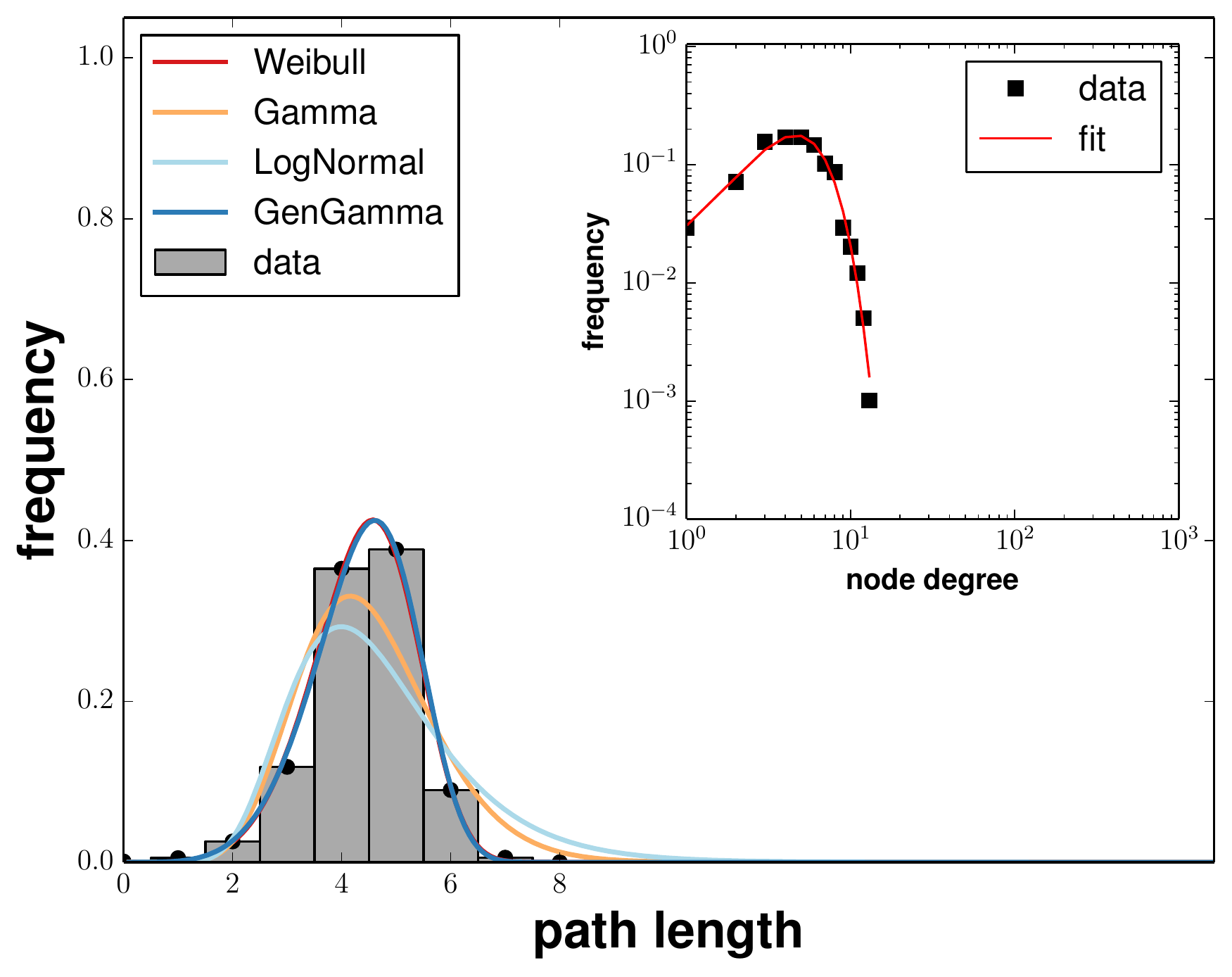}}{\ }
\subfloat[$p = 0.0075$]{\includegraphics[width=0.48\columnwidth]{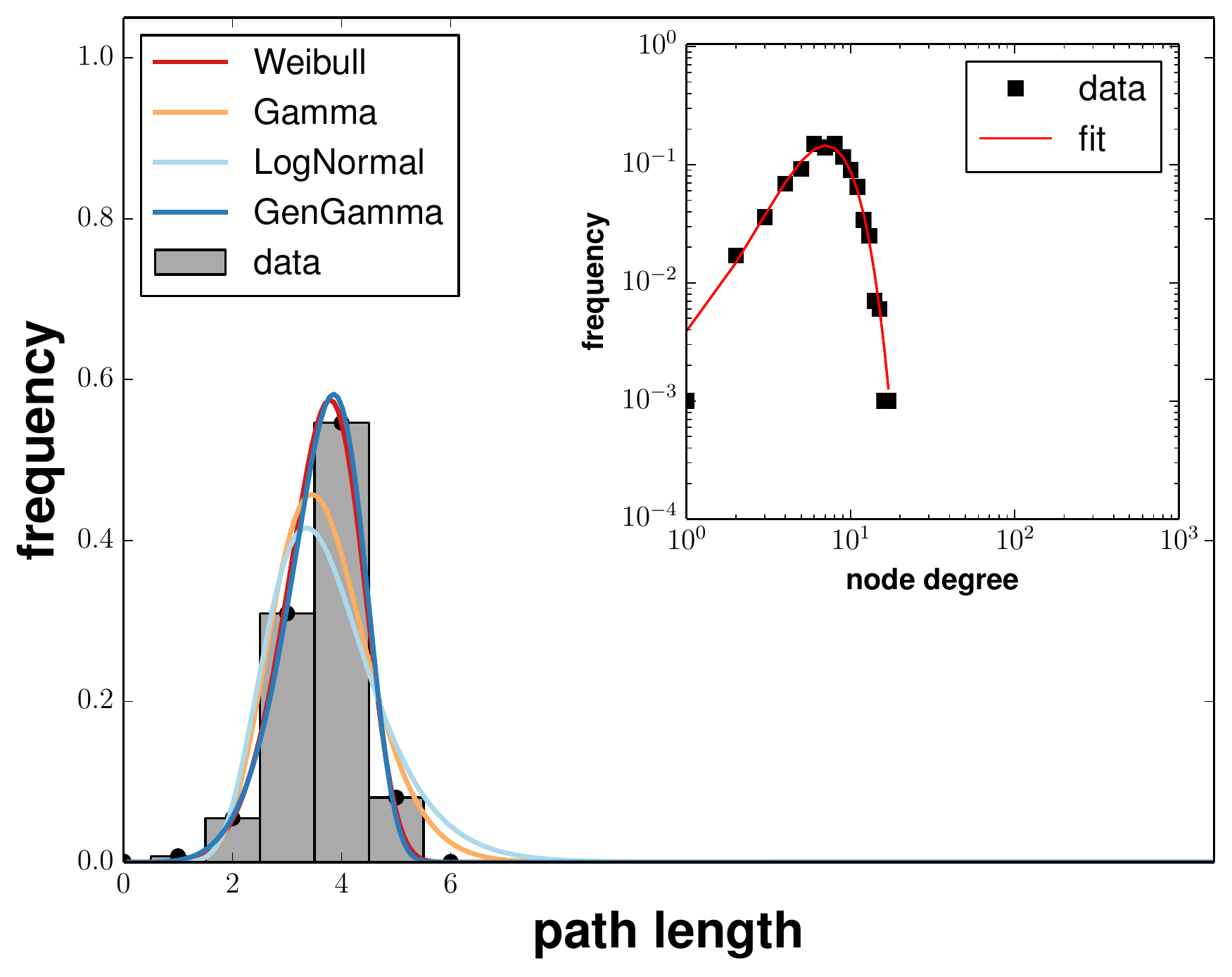}}{\ }
\caption{\label{fig:ERfits} Qualitative examples of shortest path distributions in Erd\H{o}s-R\'enyi graphs. Confirming \cite{Bauckhage2013-TWA}, the Weibull distribution fits well. Yet, the generalized Gamma distribution fits better.}
\end{figure}

Finally, Bild et al.~\cite{Bild2014-ACO} recently investigated information cascades and retweet networks on twitter and found that they could accurately fit their data using the \textit{LogNormal distribution} 
\begin{equation}
\label{eq:LN}
f_{LN}(t \mid \mu, \xi) = \frac{1}{\sqrt{2 \pi} \, \xi \, t} \, e^{-\frac{(\log t - \mu)^2}{2 \xi^2}}
\end{equation}
where $\mu$ and $\xi$ are location and scale parameters.

Regarding our approach in this paper, we emphasize that \eqref{eq:LN} was not found theoretically but emerged empirically. At the same time, we note that the Gamma, Weibull, and LogNormal distribution may easily be confused for one another \cite{Rinne2008-TWD}. Our discussion so far thus begs the question if the above results are contradictory or if they may be unified within a more general framework? Next, we show that the latter is indeed the case.

\subsection{New Result}

In this subsection, we establish the \textit{generalized Gamma distribution} as a comprehensive model of the shortest path distributions and outbreak data in connected networks of arbitrary topology.

\begin{figure}[t!]
\centering
\subfloat[$\mu = 1, \xi = 0.75$]{\includegraphics[width=0.48\columnwidth]{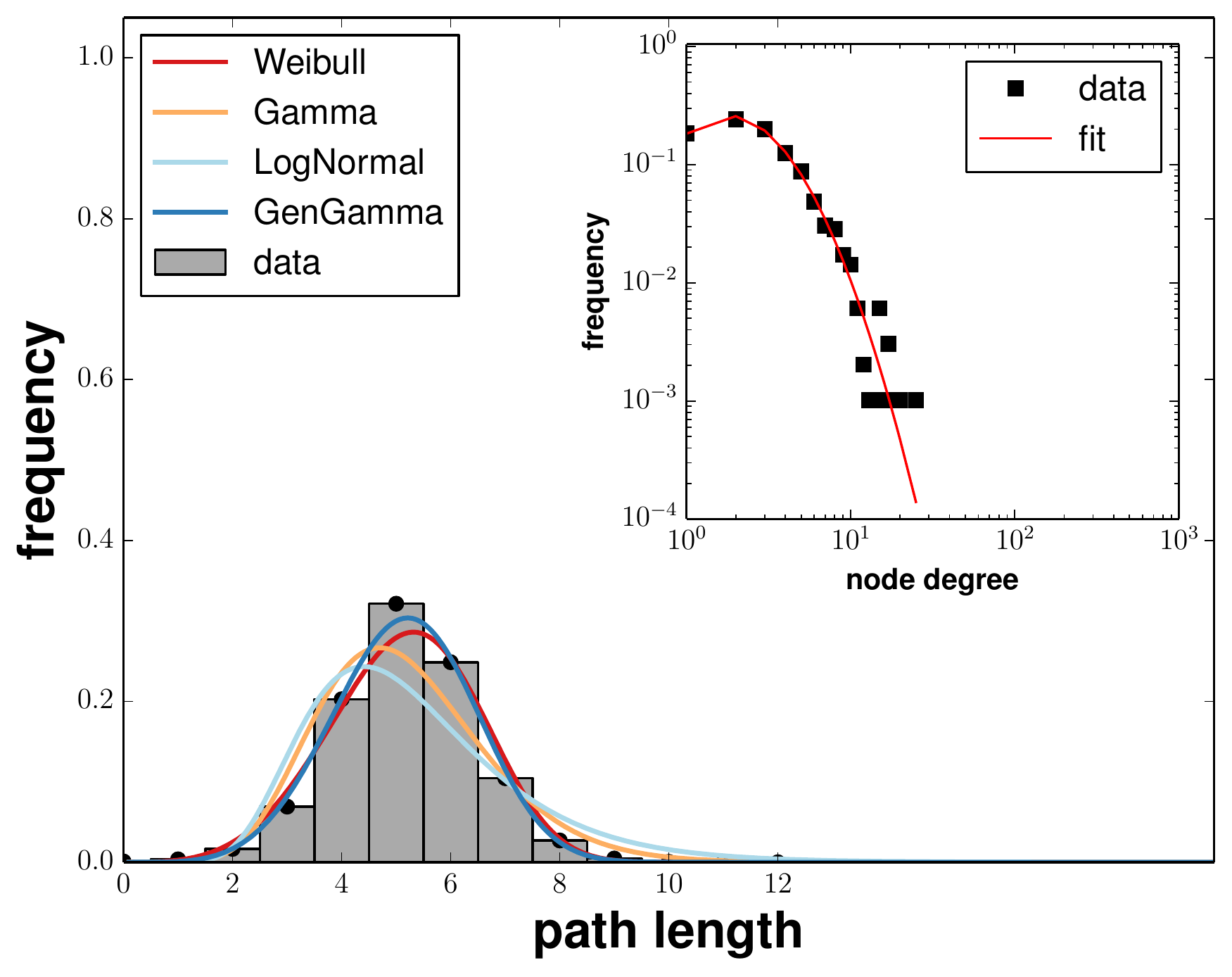}}{\ }
\subfloat[$\mu = 2, \xi = 0.75$]{\includegraphics[width=0.48\columnwidth]{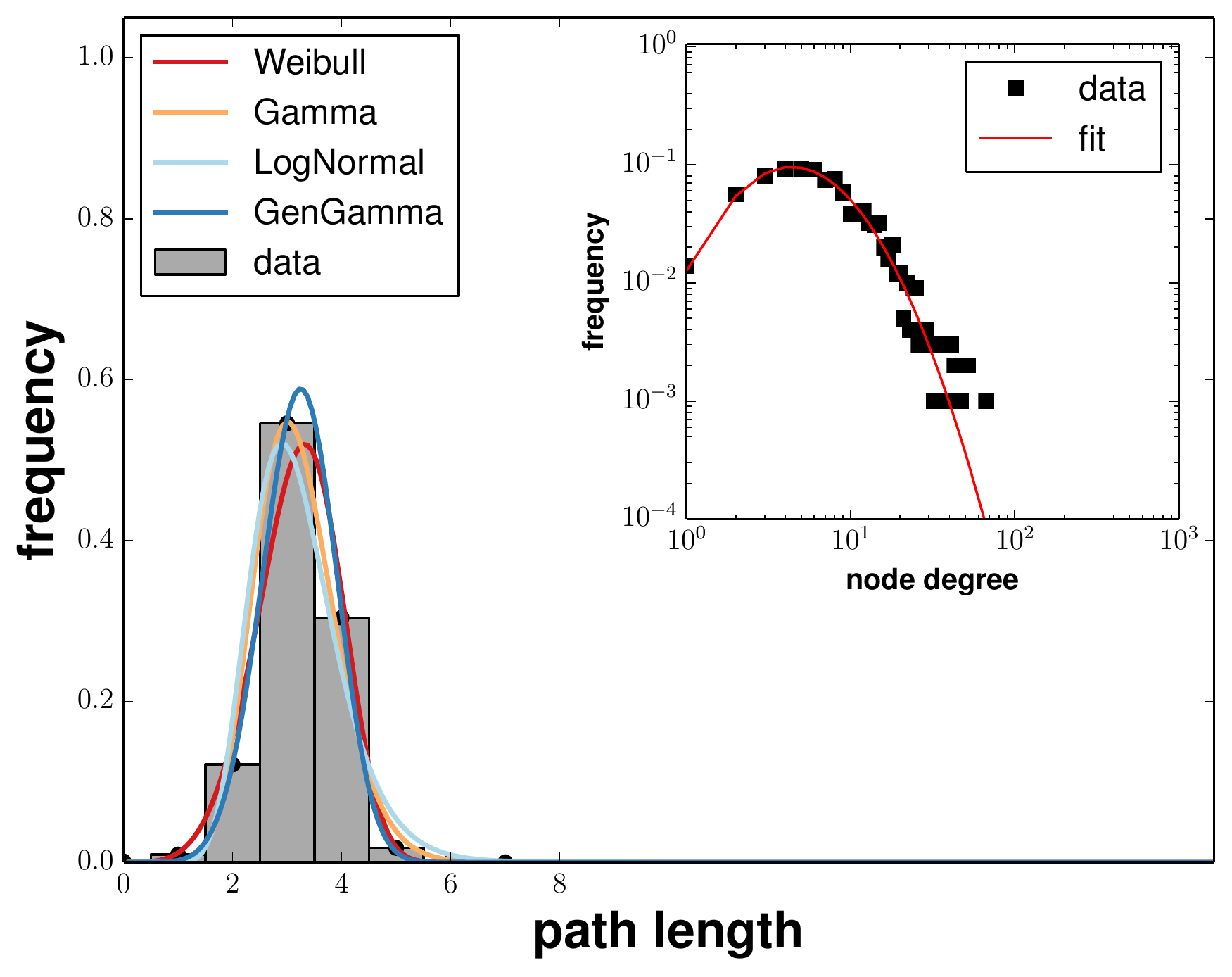}}
\caption{\label{fig:LNfits} Qualitative examples of shortest path distributions in LogNormal graphs. For different node degree distributions, the models considered here provide more or less accurate fits. The generalized Gamma, however, always fits best (cf.~Tab.~\ref{tab:synnets}).}
\end{figure}

Different versions of the generalized Gamma distribution can be traced back to 1920s \cite{Crooks2010-TAD}. Here, we are concerned with the three-parameter version that was introduced by Stacy \cite{Stacy1962-AGO}. Its probability density function is defined for $t \in [0,\infty)$ and given by
\begin{equation}
\label{eq:GG}
f_{GG}(t \mid \ggscl, \ggshpa, \ggshpb) = \frac{\ggshpb}{\ggscl^\ggshpa} \frac{1}{\Gamma(\ggshpa / \ggshpb)} t^{\ggshpa-1} e^{-(t/\ggscl)^\ggshpb}
\end{equation}
where $\ggscl > 0$ determines scale and $\ggshpa > 0$ and $\ggshpb > 0$ are shape parameters. 

The probability density function in \eqref{eq:GG} is unimodal but may be skewed to the left or to the right. It also contains several other distributions as special cases \cite{Bauckhage2014-CTK,Crooks2010-TAD}. Most notably, we point out that
\begin{itemize}
\item setting $\ggshpb=1$ yields the Gamma distribution in \eqref{eq:GA}
\item equating $\ggshpa=\ggshpb$ yields the Weibull distribution in \eqref{eq:WB}
\item letting $\ggshpa / \ggshpb \rightarrow \infty$, the generalized Gamma distribution approaches the LogNormal distribution in \eqref{eq:LN}
\end{itemize}

These properties immediately suggest that the above results might be specific instances of a more general model. Indeed, our theoretical discussion in the appendix shows that this is not coincidental but that the generalized Gamma distribution actually provides a \emph{physically plausible} model of empirical shortest path distributions and outbreak data.

In contrast to the work in \cite{Vazquez2006-PGI} and \cite{Bauckhage2013-TWA}, our derivation in appendix~\ref{sec:derivation} does not need assumptions as to the topology of the network through which a viral agent is spreading. Rather, it considers general properties of highly infectious processes as discussed in the introduction and resorts to the maximum entropy principle together with likelihood maximization techniques. Any aspects related to topological properties (e.g.~degree distribution or clustering patterns) of individual networks are absorbed into the parameters of the model in \eqref{eq:GG} which, as we shall see next, is flexible enough to represent a wide range of different path length and spreading statistics.

\section{Empirical Evaluation}
\label{sec:experiments}

In this section, we empirically evaluate the merits of the theoretical models discussed above. First, we present results on shortest path length distributions in synthesized networks. Then, we report on experiments with a large number of spreading processes on synthetic graphs and, finally, we discuss result obtained for large, real-world networks. 

Throughout, we fit continuous distributions to discrete histograms. That is, we apply functions $f(t; \vec{\theta})$ to represent counts $n_0, \ldots, n_K$ which are grouped into $K$ distinct intervals $(t_0,t_1]$, $(t_1,t_2]$, $\ldots$, $(t_{K-1}, t_\infty)$. For this setting, it is advantageous to use multinomial likelihood estimation based on reweighted least squares in order to determine optimal model parameters $\vec{\theta}^*$ \cite{Jennrich1975-MLE}. For a recent detailed exposition of this robust technique, we refer to \cite{Bauckhage2014-SRI}.

For fitted models, we report quantitative goodness-of-fit results in terms of the Hellinger distance 
\begin{equation}
\label{eq:HL}
H \bigl( h[t], f[t] \bigr) = \frac{1}{2}  \sqrt{\sum_{t} \left( \sqrt{h[t]} - \sqrt{f[t]} \right)^2}
\end{equation}
between discrete empirical data $h[t]$ and a discretized model $f[t]$ where
\begin{equation}
f[t] = 
\begin{cases}
F(t+\frac{1}{2}) & \text{ if } t = 0 \\
F(t+\frac{1}{2}) - F(t-\frac{1}{2}) & \text{ if } 0 < t < K \\
1 - F(t+\frac{1}{2}) & \text{ if } t = K
\end{cases}
\end{equation}
and $F(\cdot)$ is the corresponding cumulative density function. We note that the Hellinger distance is bound as $0 \leq H \leq 1$.

\subsection{Synthetic Networks}

\begin{table}[t!]
\caption{\label{tab:synnets} Goodness of Fit (avg. Hellinger distances) for shortest path histograms of synthetic networks}
\small
\centering
\begin{tabular}{llrrrr}
\toprule
network & parameters & $f_{WB}$ & $f_{GA}$ & $f_{LN}$ & $f_{GG}$ \\
\midrule
 ER & $\pi = 0.0050$ &  0.058 &  0.170 &  0.227 &  0.051 \\
    & $\pi = 0.0075$ &  0.046 &  0.164 &  0.205 &  0.041 \\
\midrule
 BA & $m = 1$ &  0.039 &  0.066 &  0.135 &  0.011 \\
    & $m = 2$ &  0.017 &  0.125 &  0.170 &  0.015 \\
    & $m = 3$ &  0.015 &  0.128 &  0.171 &  0.015 \\
\midrule
 PL & $\gamma = 2.1$ &  0.154 &  0.042 &  0.040 &  0.030 \\
    & $\gamma = 2.3$ &  0.132 &  0.032 &  0.046 &  0.024 \\
    & $\gamma = 2.5$ &  0.123 &  0.037 &  0.064 &  0.028 \\
    & $\gamma = 2.7$ &  0.101 &  0.044 &  0.089 &  0.028 \\
    & $\gamma = 2.9$ &  0.081 &  0.050 &  0.108 &  0.026 \\
    & $\gamma = 3.1$ &  0.070 &  0.093 &  0.137 &  0.051 \\
\midrule
 LN & $\mu = 1, \xi = 0.75$ &  0.075 &  0.093 &  0.149 &  0.037 \\
    & $\mu = 1, \xi = 1.25$ &  0.082 &  0.067 &  0.113 &  0.029 \\
    & $\mu = 2, \xi = 0.75$ &  0.073 &  0.092 &  0.145 &  0.036 \\
    & $\mu = 2, \xi = 1.25$ &  0.079 &  0.062 &  0.102 &  0.026 \\
\bottomrule
\end{tabular}
\end{table}

We created different Erd\H{o}s-R\'enyi (ER), Barab\'asi-Albert (BA), power law (PL), and LogNormal (LN) graphs of $n \in \{5,000, 10,000 \}$ nodes.

ER graphs are a staple of graph theory and merit investigation. To create ER graphs, we used edge probability parameters $\pi \in \{ 0.005, 0.0075, 0.01\}$. BA and PL graphs represent networks that result from preferential attachment processes and are frequently observed in biological, social, and technical contexts. To create BA graphs, we considered attachment parameters $m \in \{1, 2, 3\}$ and exponents of the vertex degree distributions of the PL graphs were drawn from $\gamma \in \{2.1, 2.2, \ldots, 3.2\}$. LN graphs show log-normally distributed vertex degrees and reportedly characterize link structures within sub-communities on the web \cite{Pennock2002-WDT}. To synthesize LN graphs, parameters were chosen from $\mu \in \{1, 1.5, \ldots, 3\}$ and $\xi \in \{0.25, 0.5, 0.75, 1\}$. For each parametrization of each model, we created $100$ instances.

\begin{figure}[t!]
\centering
\subfloat[$\gamma=2.3$]{\includegraphics[width=0.48\columnwidth]{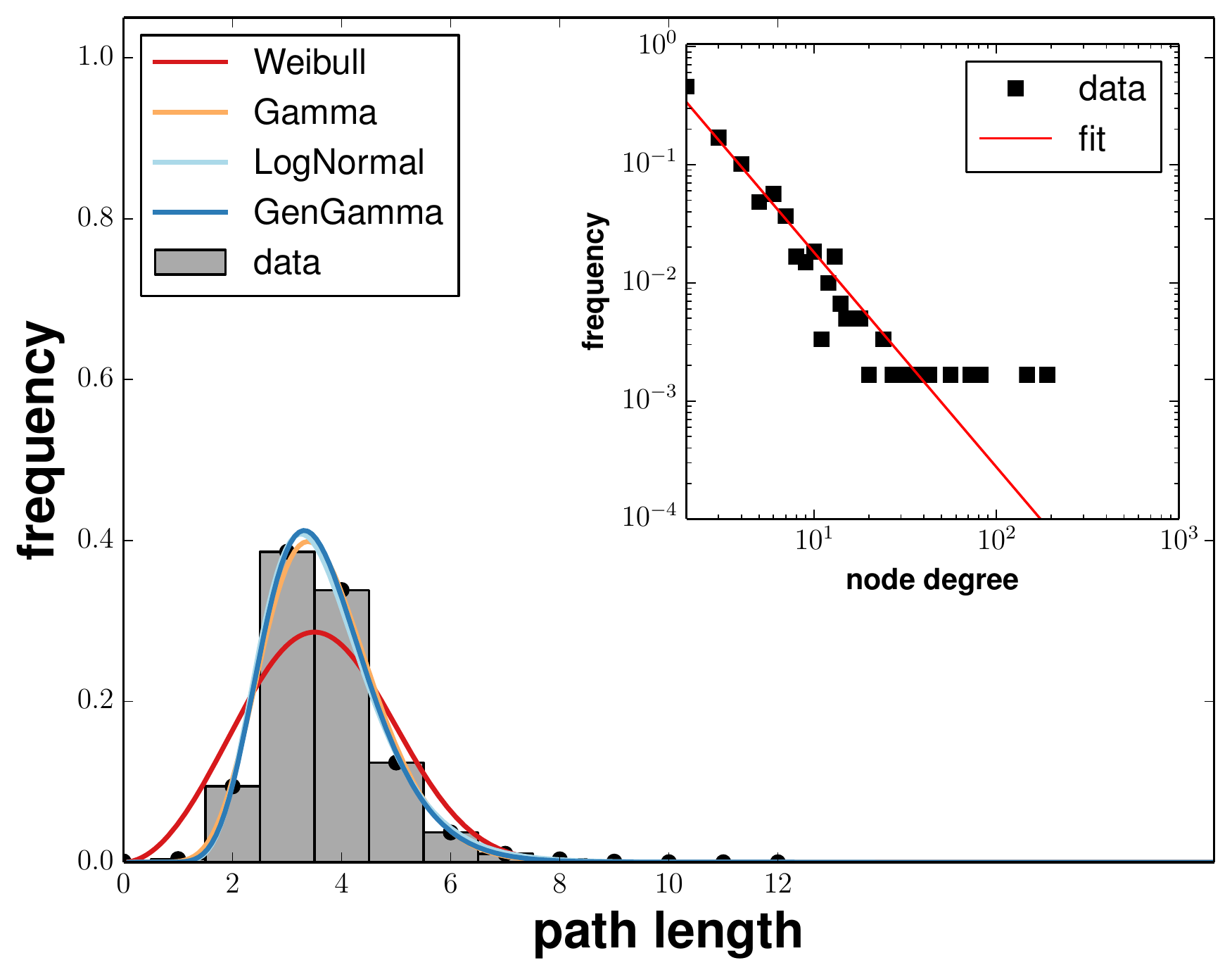}}{\ }
\subfloat[$\gamma=3.1$]{\includegraphics[width=0.48\columnwidth]{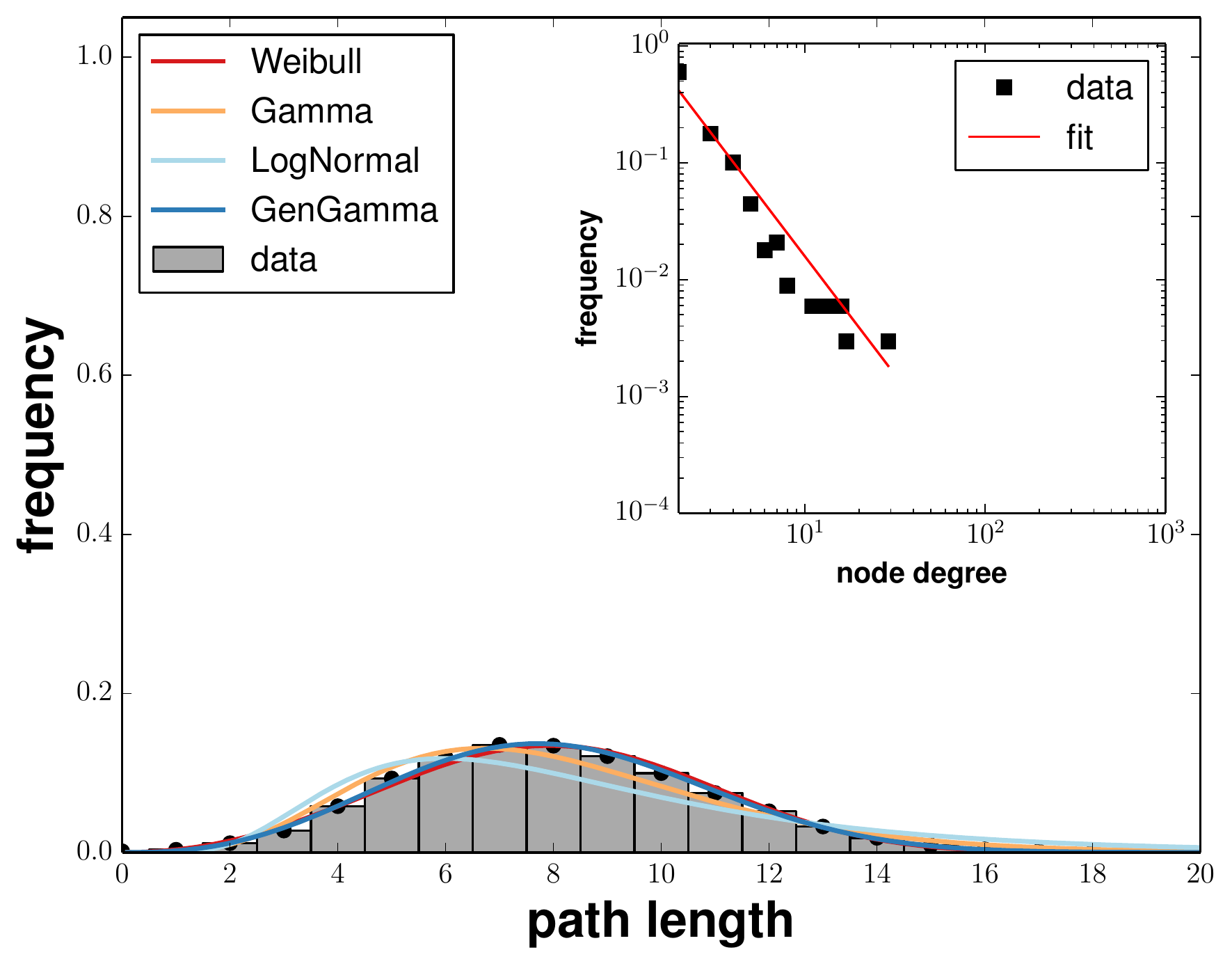}}
\caption{\label{fig:PLfits} Qualitative examples of shortest path distributions in power law graphs. Confirming \cite{Vazquez2006-PGI}, for power law coefficients $2 < \gamma < 3$, the Gamma distribution provides good fits; for $\gamma \geq 3$, the Weibull distribution provides better fits. In any case, the generalized Gamma distribution fits best.}
\end{figure}

\begin{figure}[t!]
\centering
\subfloat[$m=1$]{\includegraphics[width=0.48\columnwidth]{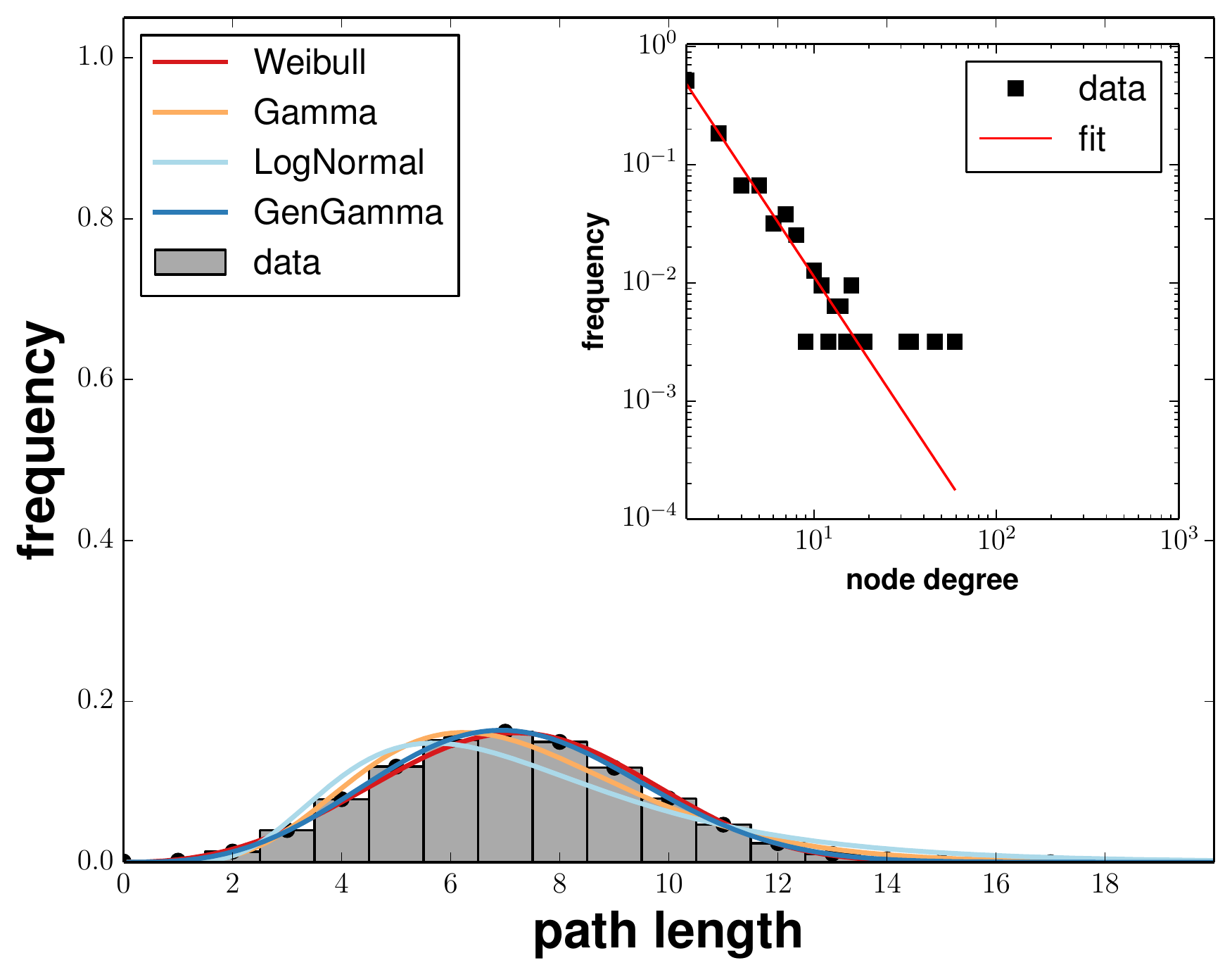}}{\ }
\subfloat[$m=3$]{\includegraphics[width=0.48\columnwidth]{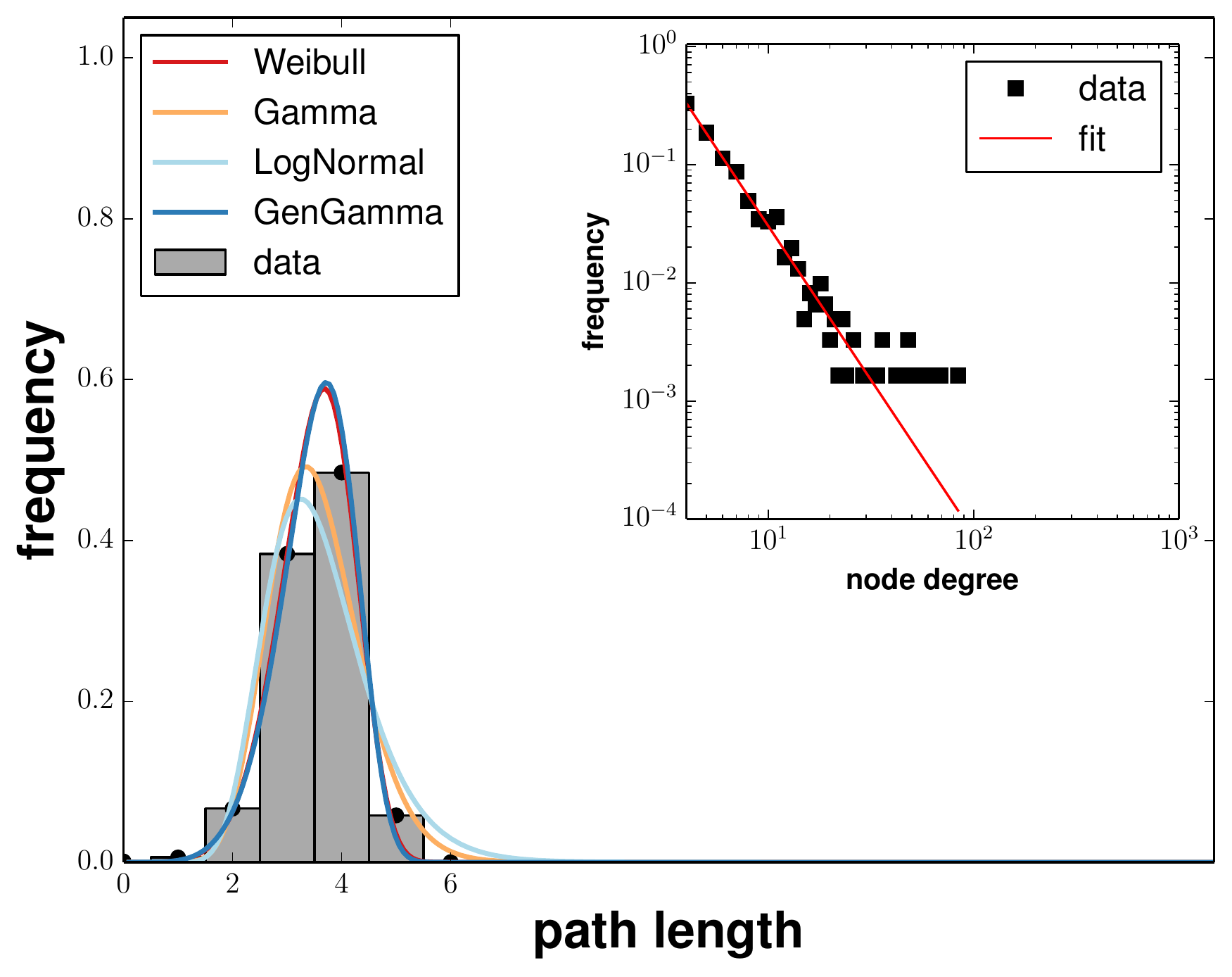}}
\caption{\label{fig:BAfits} Qualitative examples of shortest path distributions in Barab\'asi-Albert graphs. Independent of the attachment parameter $m$, node degrees are always power law distributed with $\gamma = 3$. The Weibull distribution therefore fits well, but again, the generalized Gamma distribution provides the best fit.}
\end{figure}

Examples of shortest path histograms and corresponding model fits obtained in these experiments are shown in Figs.~\ref{fig:ERfits} trough \ref{fig:BAfits}.

Table~\ref{tab:synnets} summarizes average goodness of fit results for graphs of $n = 10,000$ nodes and different topologies (for lack of space, we omit results for some of the  parametrizations considered in our experiments). We observe that \textbf{(i)} in agreement with\cite{Bauckhage2013-TWA}, the Weibull distribution provides a well fitting model for the distribution of shortest path lengths in ER graphs; it outperforms the Gamma and the LogNormal distribution; \textbf{(ii)} in agreement with \cite{Vazquez2006-PGI}, the Gamma distribution provides a well fitting model for PL graphs where $2 < \gamma < 3$; \textbf{(iii)} for PL graphs where $\gamma \lesssim 2.2$, the LogNormal fits well, too; \textbf{(iv)} for PL graphs where $\gamma \geq 3$, the Weibull fits better than the Gamma or the LogNormal; in this context, we note that BA graphs are power law graphs for which $\gamma = 3$ \cite{Barabasi1999-EOS}; \textbf{(v)} in any case, the generalized Gamma distribution provides the best fits.

The latter is hardly surprising as the densities in \eqref{eq:GA}, \eqref{eq:WB}, and \eqref{eq:LN} are functions of two parameters whereas the model in \eqref{eq:GG} depends on three parameters and therefore offers greater flexibility in statistical model fitting. Nevertheless, these results agree with the theoretical considerations above. While the Gamma and the Weibull distribution fit path length distributions in particular types of networks, the generalized Gamma distribution provides accurate fits across a wide variety of underlying network topologies.

\subsection{Spreading Processes}

\begin{table*}[t!]
\caption{\label{tab:spreads} outbreak data obtained from spreading processes in power law networks of $10,000$ nodes}
\small
\begin{tabular}{lcccccc}
\toprule
 & $i = 0.5, r = 0.5$ & $i = 0.5, r = 0.7$ & $i = 0.5, r = 0.9$ & $i = 0.9, r = 0.5$ & $i = 0.9, r = 0.7$ & $i = 0.9, r = 0.9$ \\
\midrule
\raisebox{6ex}{$\gamma = 2.2$} &
\includegraphics[width=0.13\textwidth]{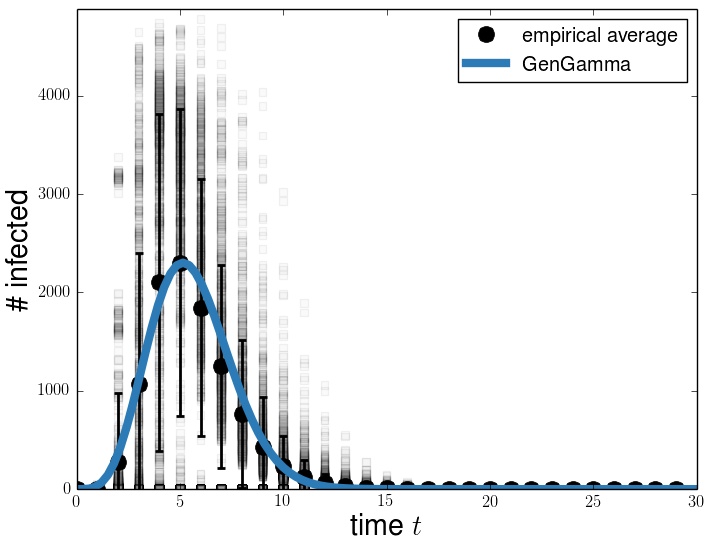} &
\includegraphics[width=0.13\textwidth]{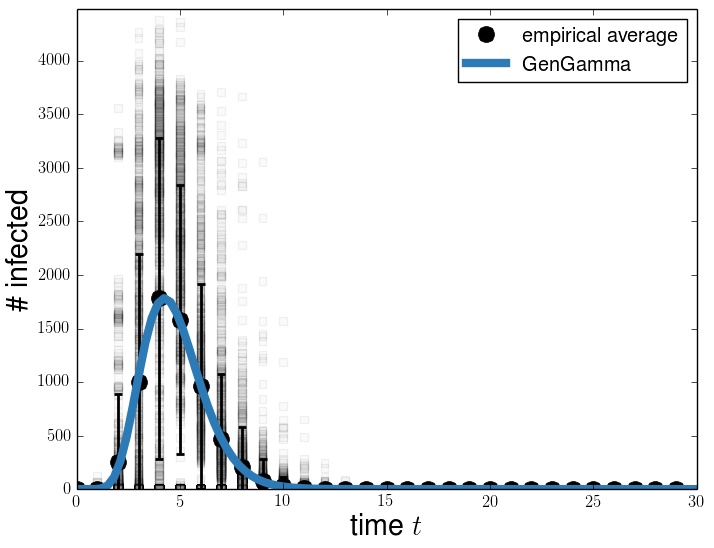} &
\includegraphics[width=0.13\textwidth]{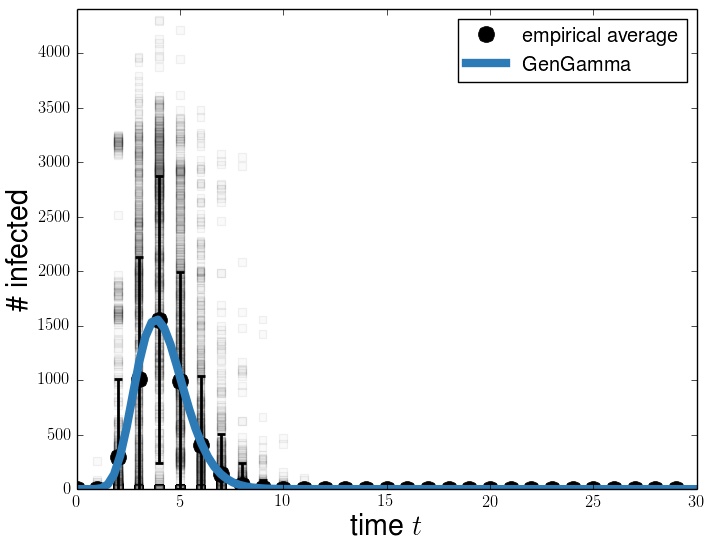} &
\includegraphics[width=0.13\textwidth]{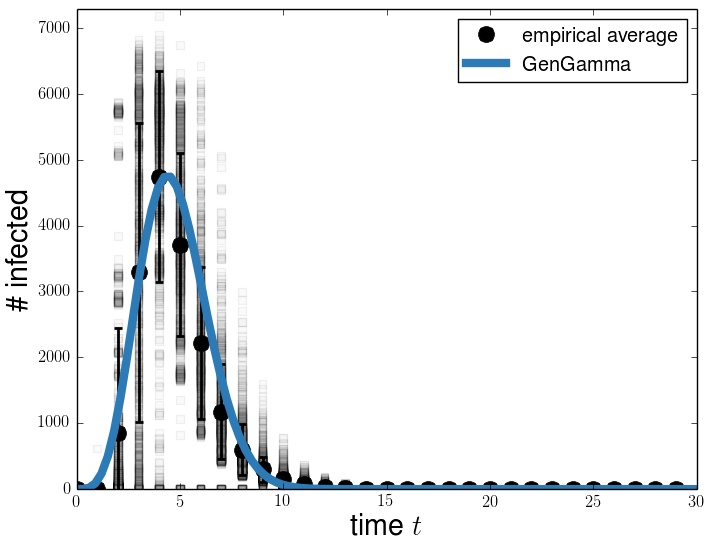} &
\includegraphics[width=0.13\textwidth]{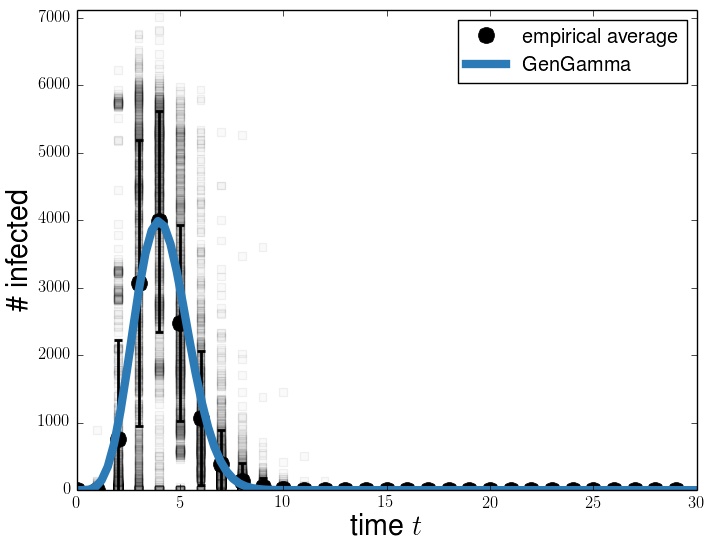}& 
\includegraphics[width=0.13\textwidth]{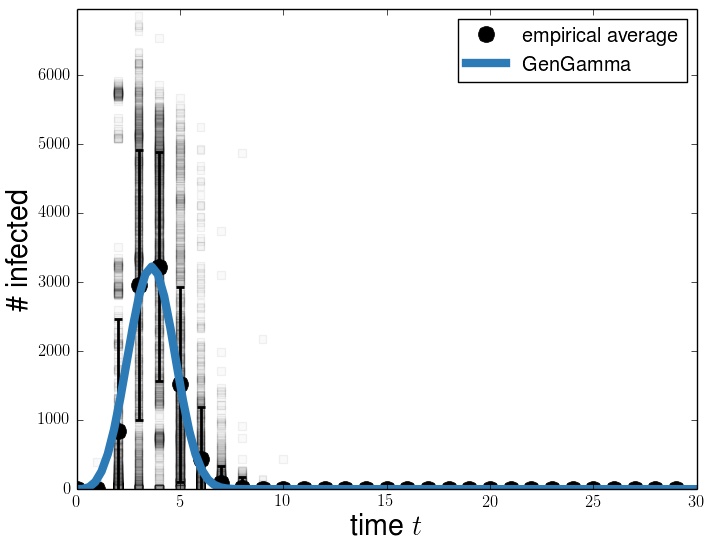} \\
\midrule
\raisebox{6ex}{$\gamma = 2.6$} &
\includegraphics[width=0.13\textwidth]{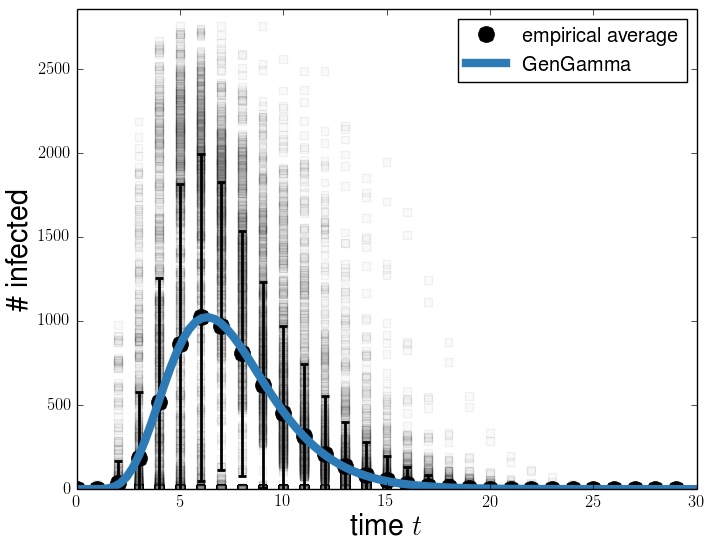} &
\includegraphics[width=0.13\textwidth]{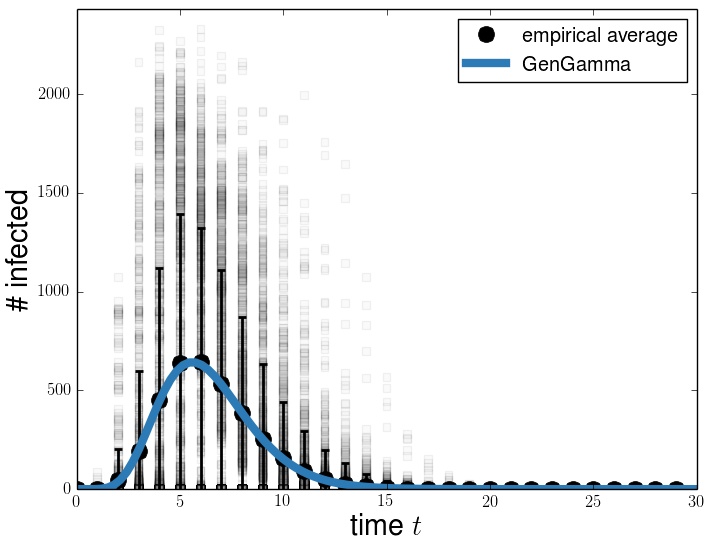} &
\includegraphics[width=0.13\textwidth]{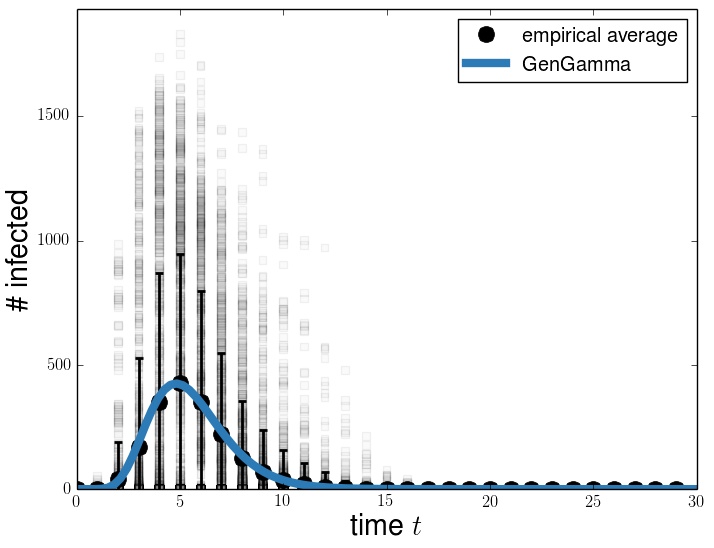} &
\includegraphics[width=0.13\textwidth]{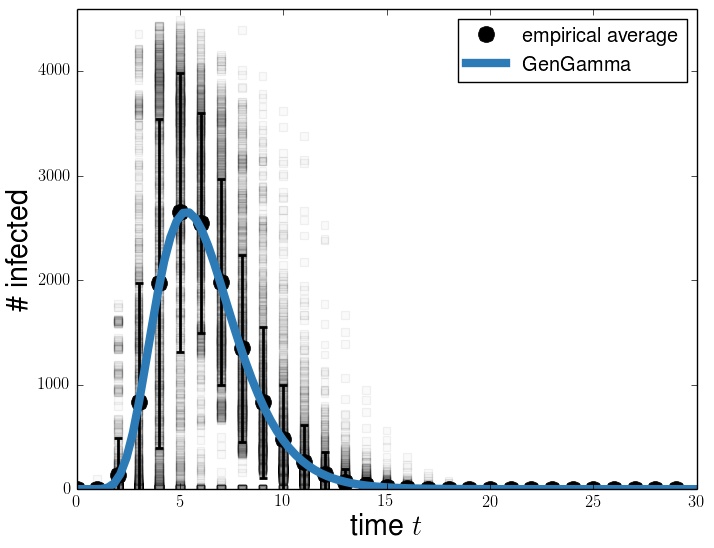} &
\includegraphics[width=0.13\textwidth]{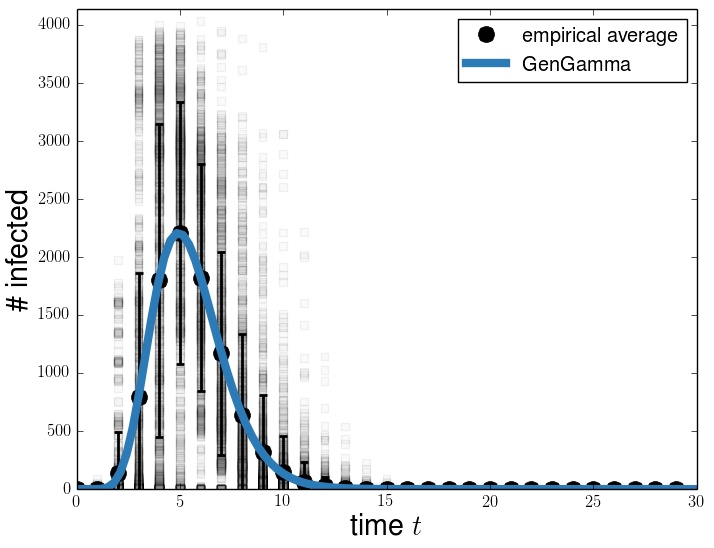}& 
\includegraphics[width=0.13\textwidth]{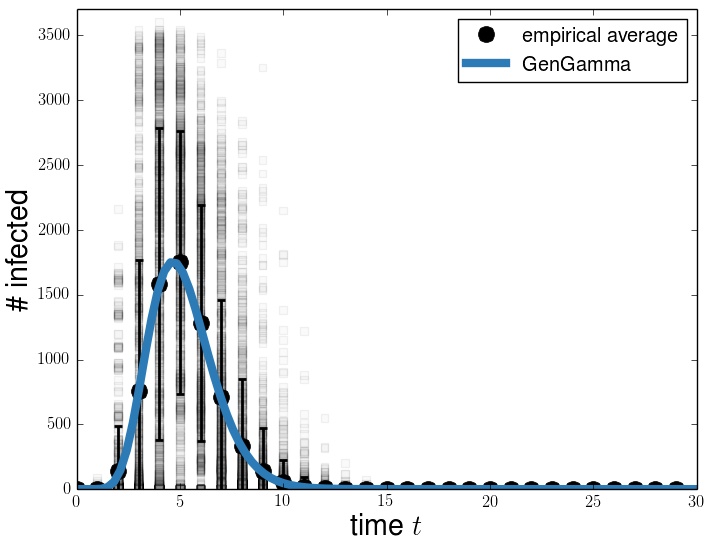} \\
\midrule
\raisebox{6ex}{$\gamma = 3.0$} &
\includegraphics[width=0.13\textwidth]{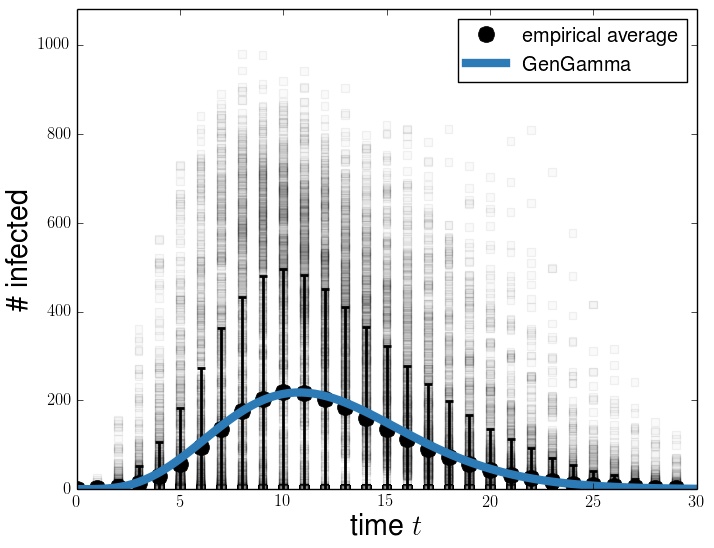} &
\includegraphics[width=0.13\textwidth]{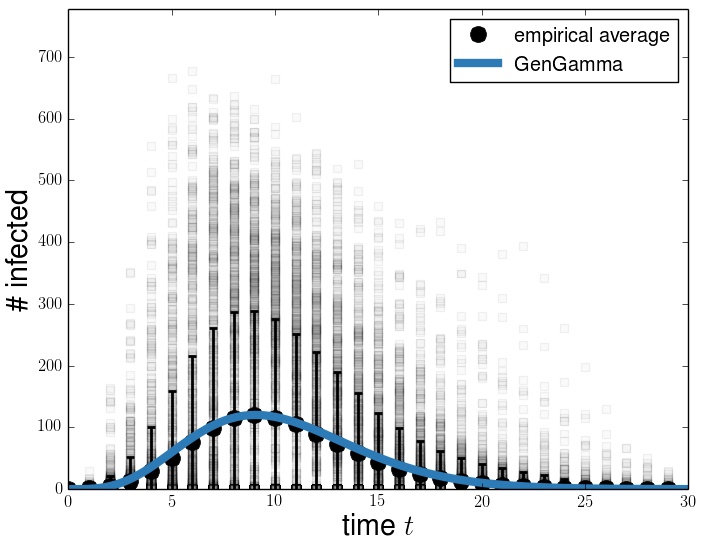} &
\includegraphics[width=0.13\textwidth]{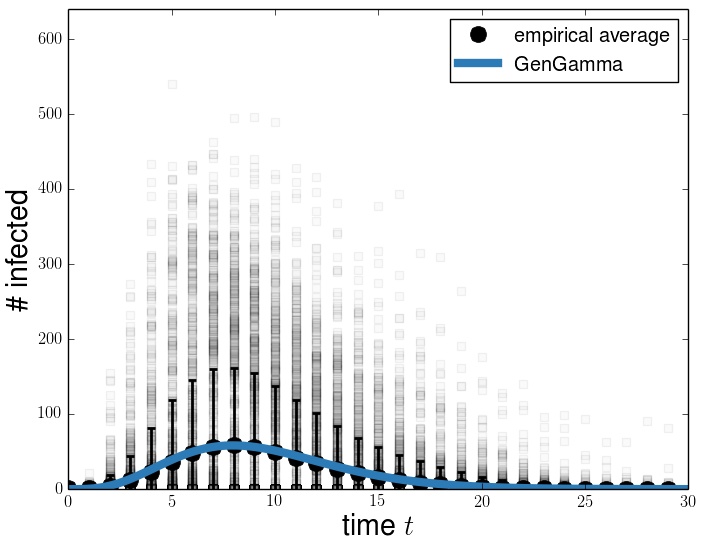} &
\includegraphics[width=0.13\textwidth]{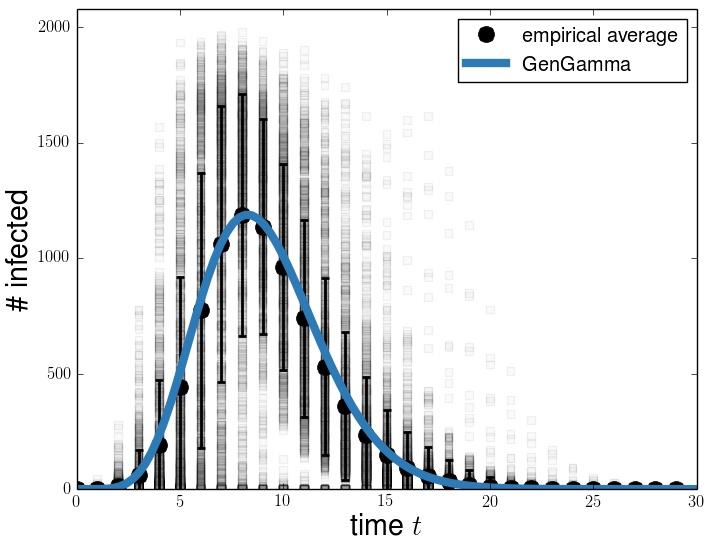} &
\includegraphics[width=0.13\textwidth]{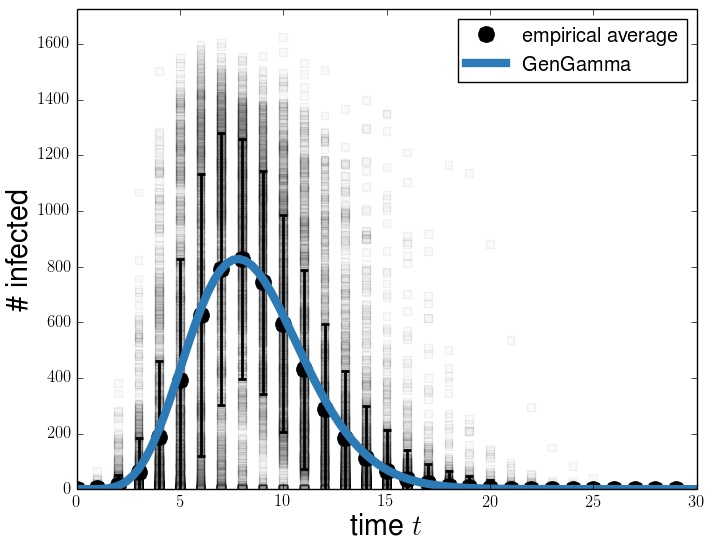}& 
\includegraphics[width=0.13\textwidth]{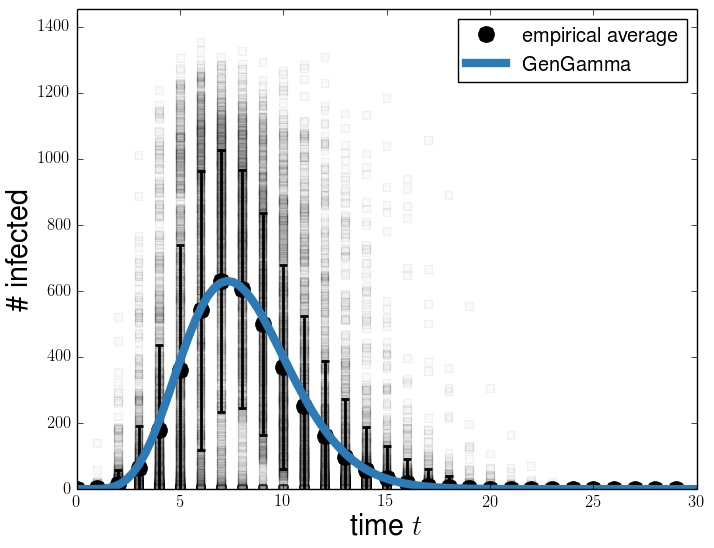} \\
\bottomrule
\end{tabular}
\end{table*}

Given the synthetic networks created above, we simulated $SIR$ spreading processes where the infection rate varied in $i \in \{0.5, 0.6, \ldots, 0.9\}$ and the recovery rate was chosen from $r \in \{0.5, 0.6, \ldots, 0.9\}$. For each network and each choice of $i$ and $r$, we created $10$ epidemics starting at randomly selected source nodes $v_s$ and fitted the generalized Gamma distribution to the resulting outbreak data.

Table~\ref{tab:spreads} shows exemplary results obtained for PL graphs of $10,000$ nodes; rows correspond to different power law exponents $\gamma$ and columns indicate different choices of pairs $(i,r)$. Each panel plots the outbreak data of all the corresponding epidemics (grey dots), the corresponding empirical average taken over the individual outbreak distributions (black dots), as well as a generalized Gamma fit to these averages (blue curves). 

Visual inspection of these results suggests that the generalized Gamma distribution accounts very well for average outbreak dynamics in power law graphs. Though not shown in the table, this behavior was also observed for individual epidemics as well as for epidemics on other networks. In appendix~\ref{sec:timescales}, we present a theoretical justification for this empirically observed behavior.

\subsection{Real-World Networks}

\begin{table}[t!]
\caption{\label{tab:socnets} Goodness of Fit (Hellinger distances) for shortest path histograms of social networks}
\small
\centering
\begin{tabular}{lrrrr}
\toprule
network & $f_{WB}$ & $f_{GA}$ & $f_{LN}$ & $f_{GG}$ \\
\midrule
advogato & 0.110 & 0.014 & 0.056 & 0.012 \\
arenas email & 0.044 & 0.036 & 0.074 & 0.008 \\
arenas pgp & 0.100 & 0.022 & 0.057 & 0.013 \\
ca AstroPh & 0.157 & 0.021 & 0.047 & 0.021 \\
ca cit HepPh & 0.133 & 0.023 & 0.065 & 0.015 \\
ca cit HepTh & 0.089 & 0.016 & 0.026 & 0.010 \\
catster & 0.101 & 0.023 & 0.032 & 0.016 \\
cfinder google & 0.076 & 0.021 & 0.023 & 0.032 \\
cit HepPh & 0.178 & 0.037 & 0.065 & 0.032 \\
cit HepTh & 0.135 & 0.018 & 0.051 & 0.016 \\
dblp cite & 0.078 & 0.042 & 0.077 & 0.012 \\
dogster & 0.215 & 0.036 & 0.048 & 0.026 \\
elec & 0.046 & 0.057 & 0.096 & 0.020 \\
email EuAll & 0.363 & 0.331 & 0.116 & 0.232 \\
enron & 0.122 & 0.046 & 0.075 & 0.023 \\
facebook wosn links & 0.186 & 0.022 & 0.037 & 0.018 \\
facebook wosn wall & 0.172 & 0.025 & 0.048 & 0.021 \\
filmtipset friend & 0.140 & 0.024 & 0.052 & 0.017 \\
gottron net all & 0.075 & 0.038 & 0.070 & 0.029 \\
gottron net core & 0.049 & 0.024 & 0.060 & 0.006 \\
hep th citations & 0.122 & 0.011 & 0.036 & 0.009 \\
loc brightkite edges & 0.193 & 0.027 & 0.035 & 0.059 \\
munmun digg reply & 0.170 & 0.028 & 0.054 & 0.023 \\
munmun twitter social & 0.132 & 0.089 & 0.079 & 0.071 \\
collaboration & 0.137 & 0.097 & 0.074 & 0.045 \\
ucsocial & 0.084 & 0.012 & 0.056 & 0.008 \\
petster carnivore & 0.120 & 0.116 & 0.119 & 0.117 \\
petster friendships cat & 0.109 & 0.023 & 0.032 & 0.027 \\
petster friendships dog & 0.206 & 0.036 & 0.048 & 0.026 \\
petster friendships hamster & 0.167 & 0.083 & 0.049 & 0.060 \\
petster hamster & 0.081 & 0.011 & 0.033 & 0.007 \\
sap & 0.142 & 0.063 & 0.090 & 0.052 \\
slashdot threads & 0.267 & 0.082 & 0.055 & 0.102 \\
slashdot zoo & 0.192 & 0.026 & 0.052 & 0.032 \\
wikiconflict & 0.134 & 0.014 & 0.055 & 0.014 \\
wikisigned k2 & 0.268 & 0.166 & 0.070 & 0.136 \\
wikisigned nontext & 0.291 & 0.086 & 0.058 & 0.066 \\
\midrule
avg. & 0.146 & 0.050 & 0.059 & 0.039 \\
\bottomrule
\end{tabular}
\end{table}

\begin{table}[t!]
\caption{\label{tab:natnets} Goodness of Fit (Hellinger distances) for shortest path histograms of natural networks}
\small
\centering
\begin{tabular}{lrrrr}
\toprule
network & $f_{WB}$ & $f_{GA}$ & $f_{LN}$ & $f_{GG}$ \\
\midrule
arenas meta & 0.090 & 0.036 & 0.025 & 0.029 \\
as caida20071105 & 0.277 & 0.160 & 0.050 & 0.113 \\
as20000102 & 0.065 & 0.016 & 0.054 & 0.005 \\
dbpedia similar & 0.685 & 0.686 & 0.082 & 0.018 \\
eat & 0.012 & 0.075 & 0.115 & 0.003 \\
lasagne frenchbook & 0.017 & 0.001 & 0.058 & 0.003 \\
openflights & 0.139 & 0.025 & 0.040 & 0.022 \\
powergrid & 0.745 & 0.745 & 0.150 & 0.016 \\
usairport & 0.053 & 0.010 & 0.040 & 0.007 \\
sociopatterns infectious & 0.030 & 0.026 & 0.053 & 0.012 \\
topology & 0.130 & 0.021 & 0.055 & 0.019 \\
wordnet words & 0.192 & 0.031 & 0.054 & 0.022 \\
wordnet & 0.133 & 0.188 & 0.214 & 0.131 \\
\midrule
avg. & 0.198 & 0.155 & 0.076 & 0.031 \\
\bottomrule
\end{tabular}
\end{table}

\begin{table}[t!]
\caption{\label{tab:bipnets} Goodness of Fit (Hellinger distances) for shortest path histograms of bipartite networks}
\small
\centering
\begin{tabular}{lrrrr}
\toprule
network & $f_{WB}$ & $f_{GA}$ & $f_{LN}$ & $f_{GG}$ \\
\midrule
adjnoun & 0.028 & 0.037 & 0.065 & 0.012 \\
bx & 0.264 & 0.096 & 0.090 & 0.130 \\
dbpedia occupation & 0.165 & 0.103 & 0.113 & 0.101 \\
dbpedia producer & 0.734 & 0.734 & 0.152 & 0.156 \\
dbpedia starring & 0.721 & 0.721 & 0.034 & 0.032 \\
dbpedia writer & 0.754 & 0.754 & 0.096 & 0.072 \\
epinions & 0.209 & 0.036 & 0.041 & 0.026 \\
escorts & 0.097 & 0.048 & 0.076 & 0.031 \\
filmtipset comment & 0.060 & 0.049 & 0.089 & 0.007 \\
github & 0.186 & 0.078 & 0.082 & 0.078 \\
gottron reuters & 0.128 & 0.112 & 0.137 & 0.106 \\
movielens 10m ti & 0.187 & 0.082 & 0.101 & 0.074 \\
movielens 10m ui & 0.036 & 0.083 & 0.116 & 0.031 \\
movielens 10m ut & 0.199 & 0.146 & 0.159 & 0.144 \\
movielens 1m & 0.043 & 0.005 & 0.040 & 0.007 \\
ucforum & 0.047 & 0.050 & 0.080 & 0.030 \\
pics ut & 0.268 & 0.164 & 0.166 & 0.183 \\
prosper support & 0.230 & 0.259 & 0.232 & 0.208 \\
youtube groupmemberships & 0.192 & 0.115 & 0.119 & 0.115 \\
\midrule
avg. & 0.239 & 0.193 & 0.105 & 0.081 \\
\bottomrule
\end{tabular}
\end{table}

The KONECT network collection \cite{Kunegis2013-KON} provides a comprehensive set of large scale, real-world network data freely available for research. Networks contained in this collection comprise (online) social networks where edges indicate social contacts or friendship relations, natural networks such as power grids or connections between airports, and bipartite networks such as typically found in the context of recommender systems. The sizes of these networks vary between $O(10,000)$ to $O(1,000,000)$ nodes and they show different node degree distributions and clustering coefficients. For further details, we refer to \href{http://konect.uni-koblenz.de/}{http://konect.uni-koblenz.de/}.

Tables~\ref{tab:socnets} through \ref{tab:bipnets} summarize goodness-of-fit results obtained from fitting the models in \eqref{eq:GA}, \eqref{eq:WB}, \eqref{eq:LN}, and \eqref{eq:GG} to hop count distributions of social, natural, and bipartite networks respectively. Again in agreement with the theoretical prediction in \cite{Vazquez2006-PGI}, we observe that the Gamma distribution provides accurate fits to path length distributions in social networks which are often reported to be power law networks with power law exponents $2 < \gamma < 3$.

\begin{figure*}[t!]
\centering
\subfloat[facebook wall post network]{\includegraphics[width=0.32\textwidth]{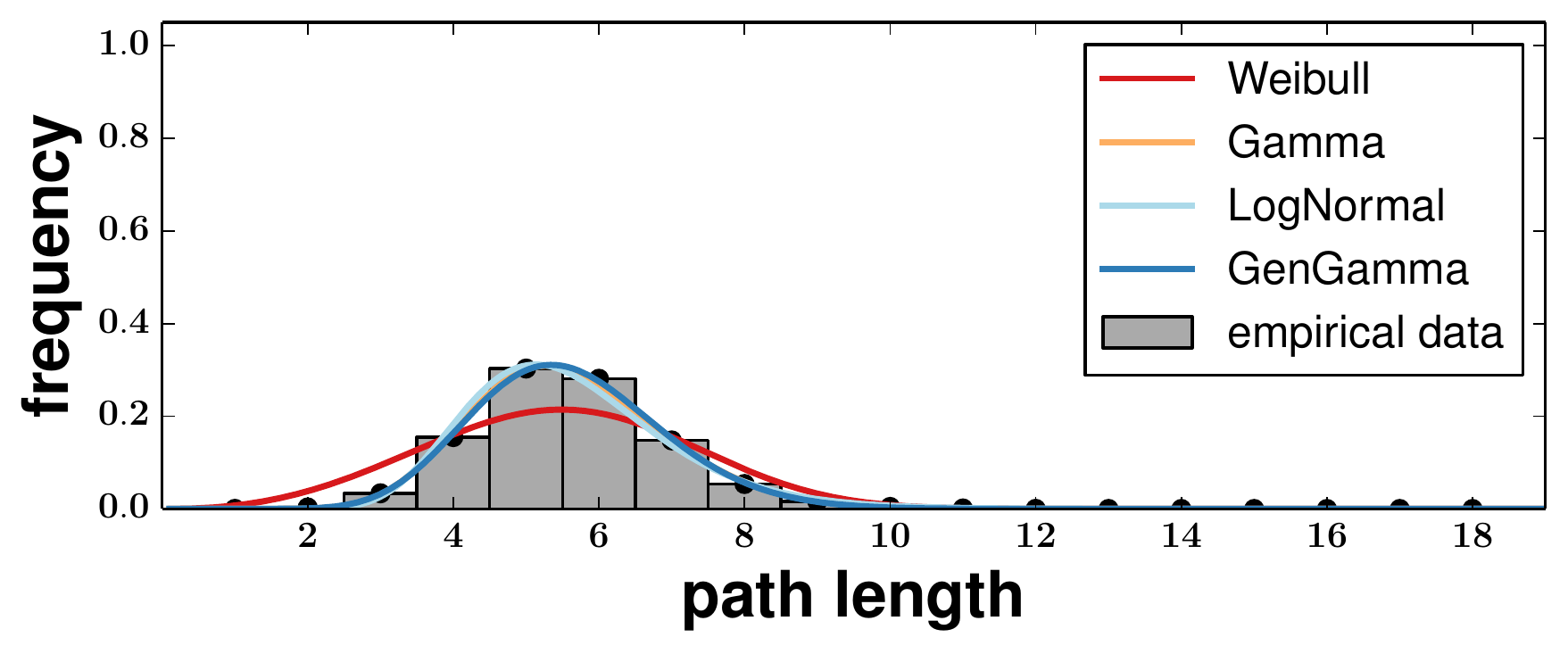}}{\ }
\subfloat[openflights]{\includegraphics[width=0.32\textwidth]{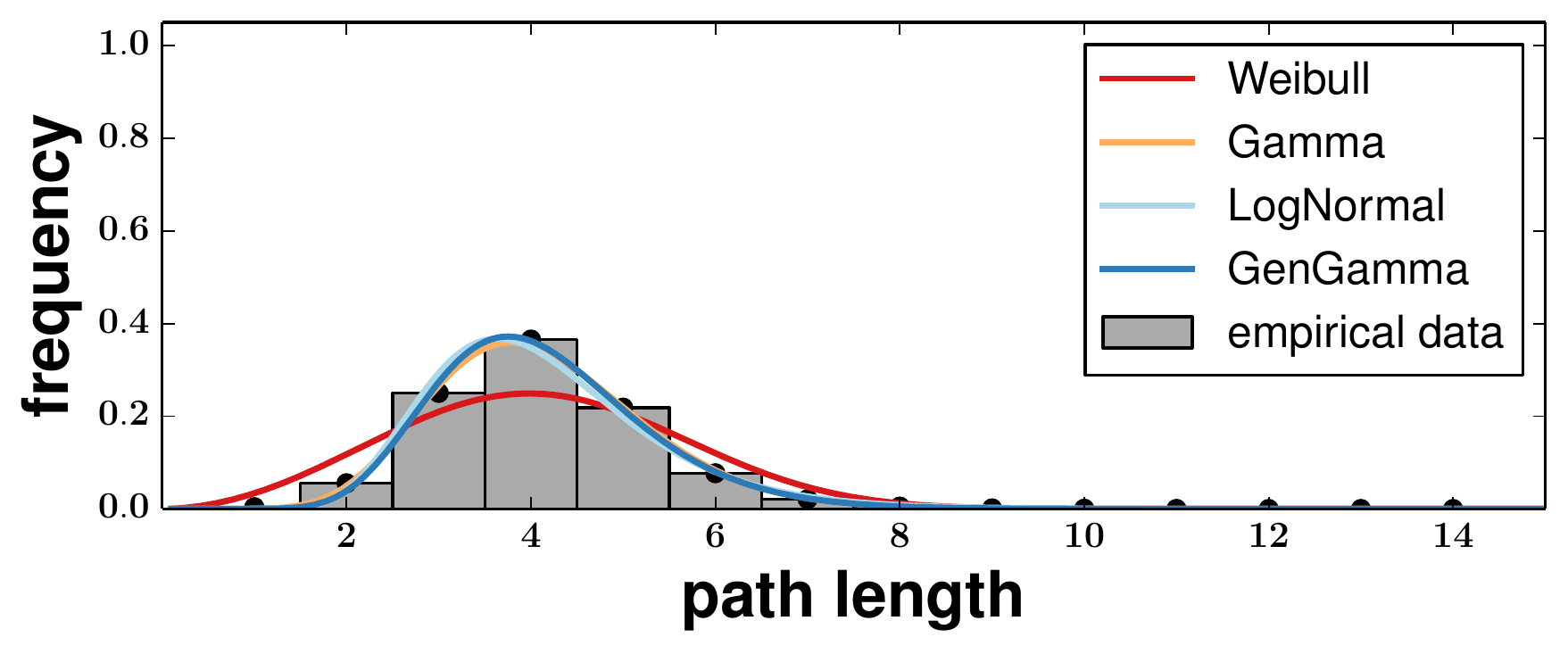}}{\ }
\subfloat[movielens usr mov]{\includegraphics[width=0.32\textwidth]{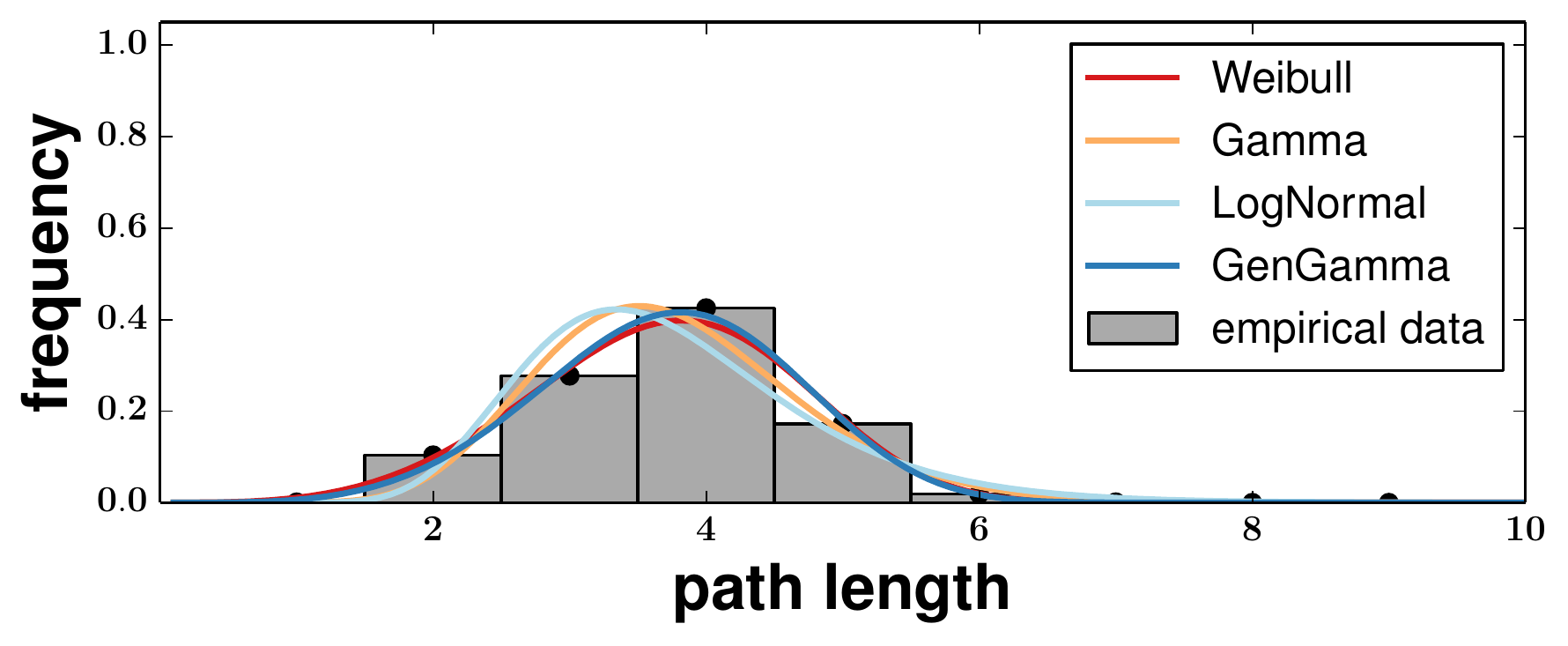}}{\ }
\caption{\label{fig:fits} Qualitative examples of shortest path distributions in real-world networks. Path length distributions of social and natural networks are well accounted for by the Gamma distribution whereas the Weibull distribution provides better fits for bipartite networks that occur in recommender settings. In each case, however, the generalized Gamma distribution accounts best for the empirical data.}
\end{figure*}

\begin{figure}[t]
\centering
\subfloat[$f_{WB}$]{\includegraphics[width=0.45\columnwidth]{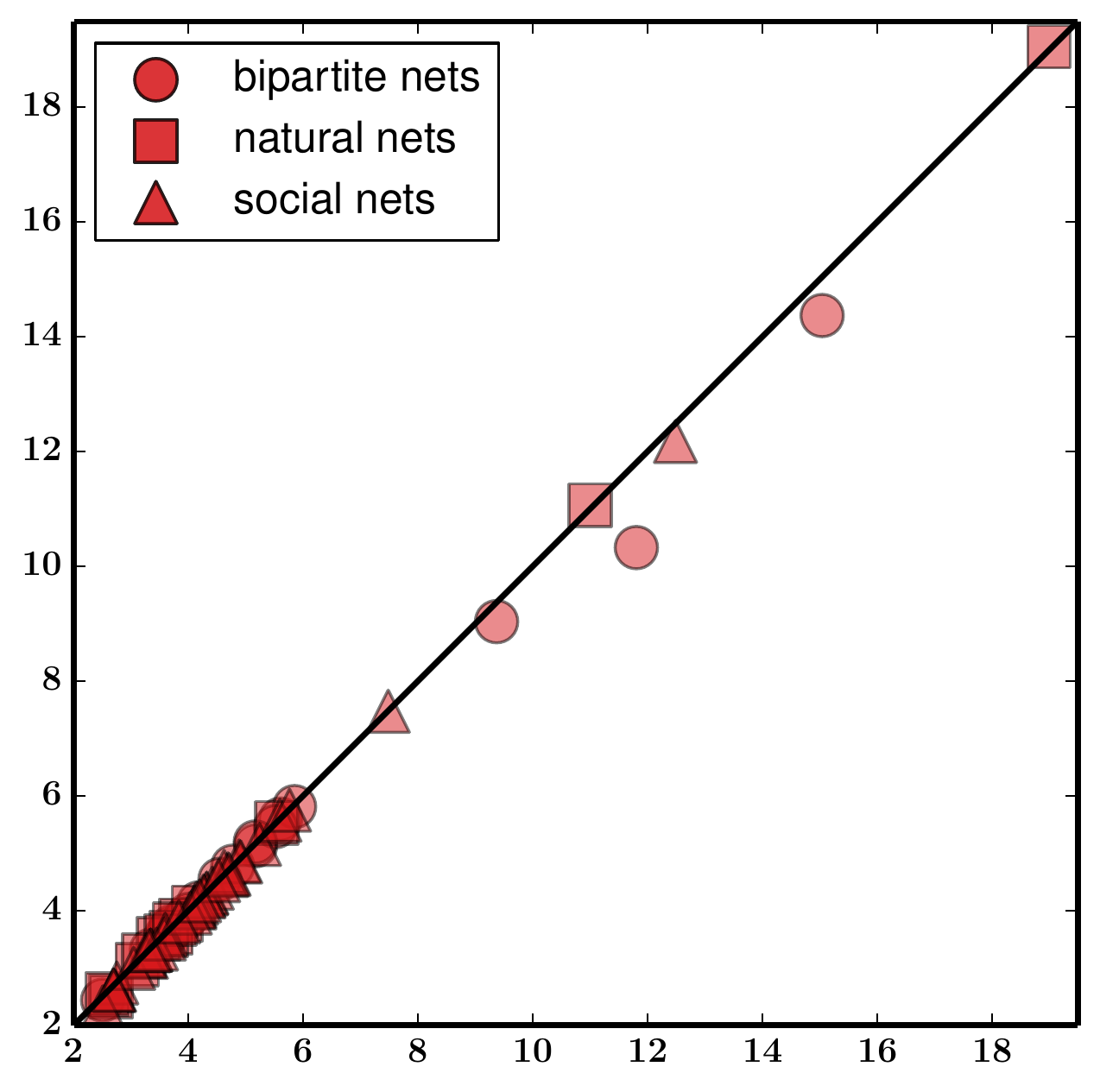}}{\ }
\subfloat[$f_{GA}$]{\includegraphics[width=0.45\columnwidth]{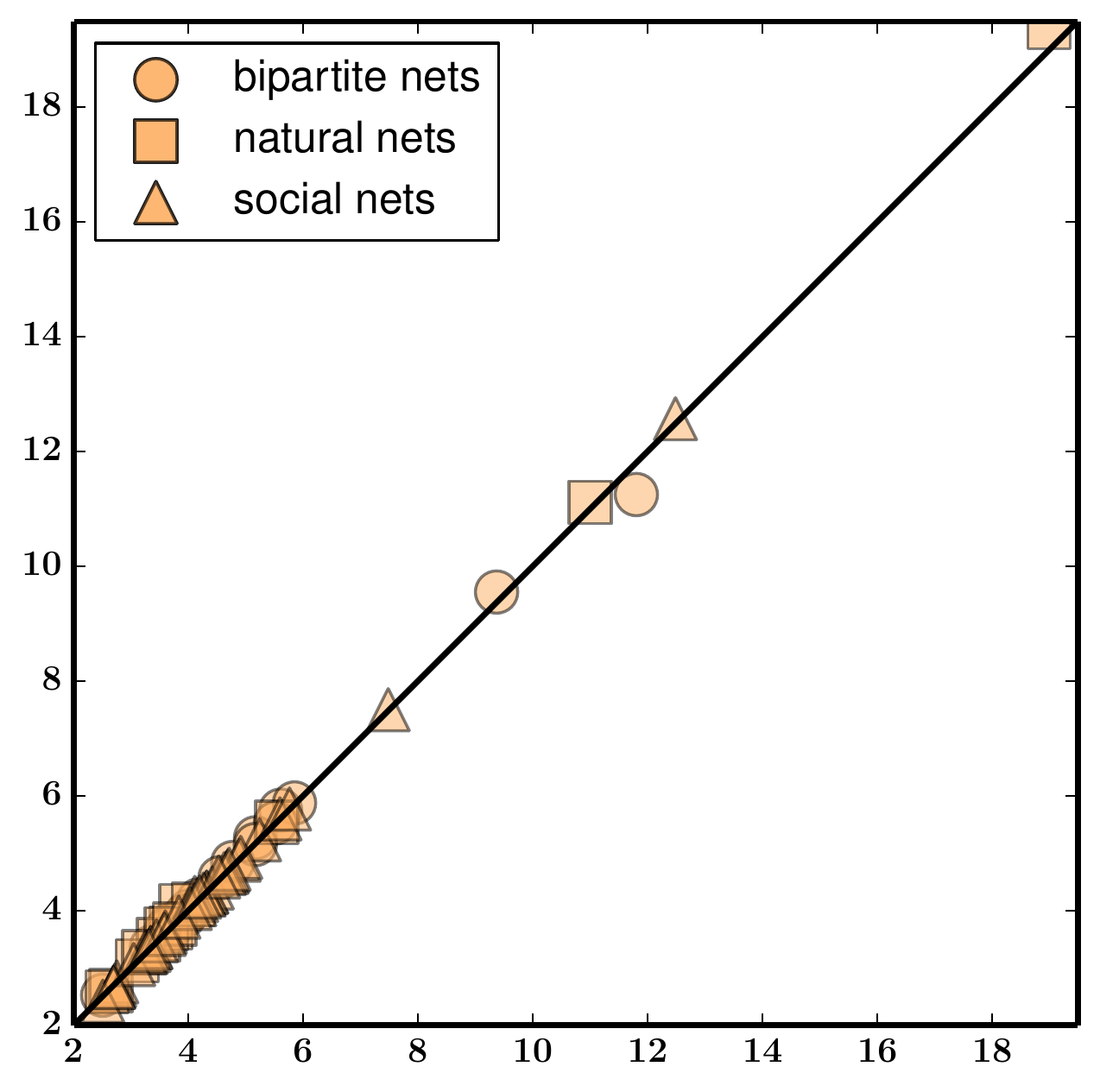}}

\subfloat[$f_{LN}$]{\includegraphics[width=0.45\columnwidth]{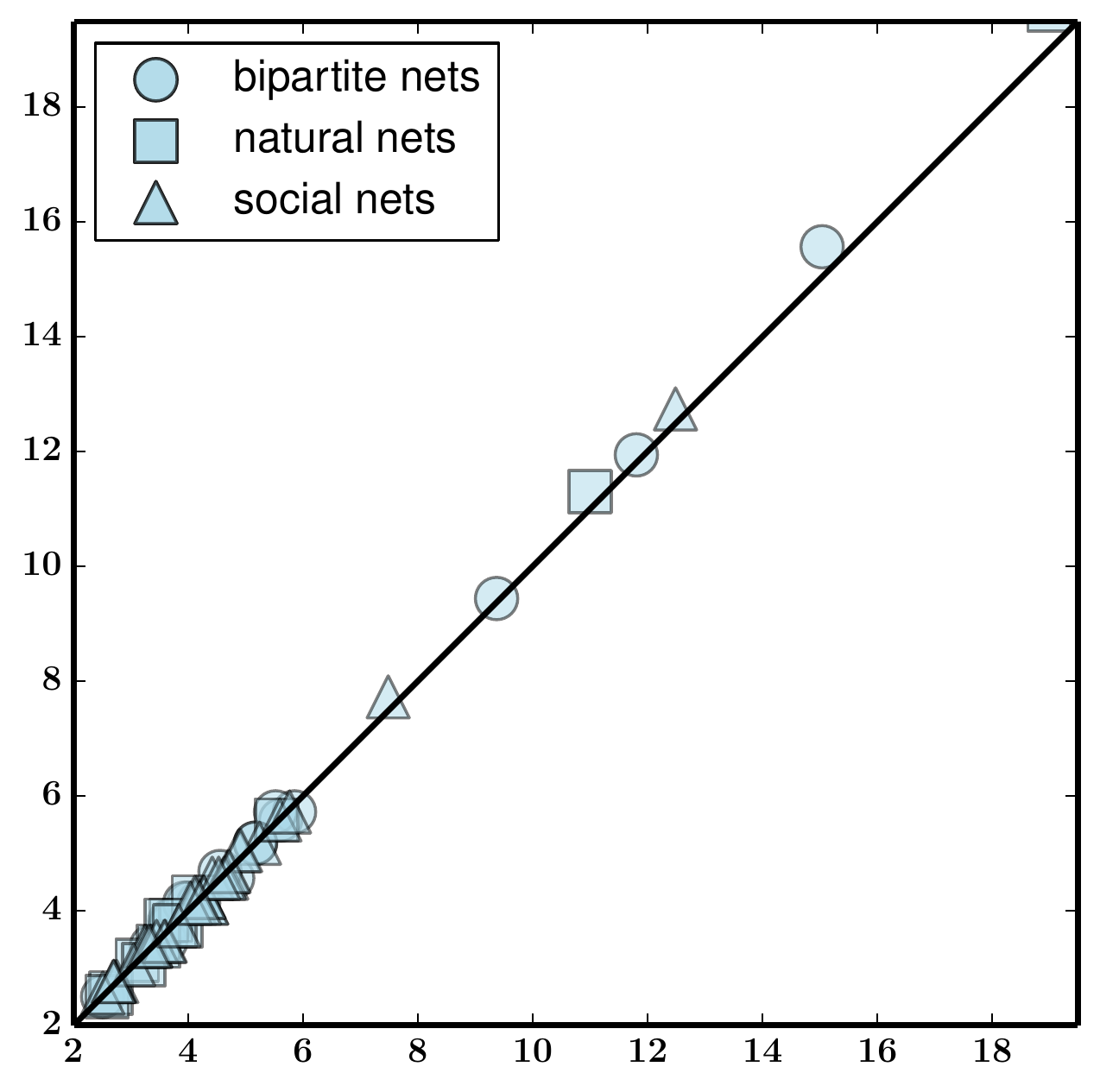}}{\ }
\subfloat[$f_{GG}$]{\includegraphics[width=0.45\columnwidth]{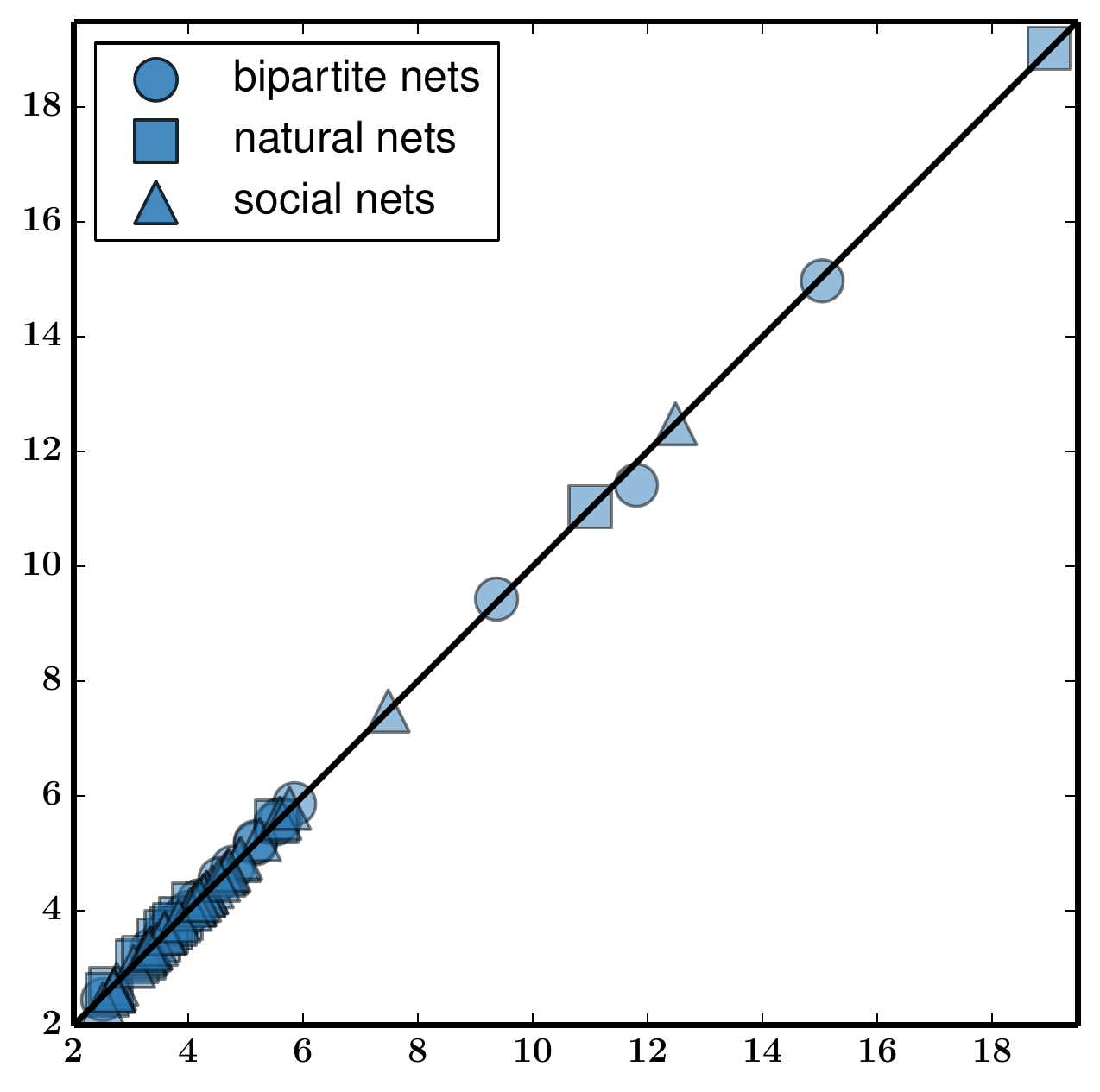}}
\caption{\label{fig:means} Empirical vs.~predicted average shortest path lengths. Predictions correspond to the means of Weibull, Gamma, LogNormal, and generalized Gamma distributions that were fitted to shortest path lengths histograms of the real world networks in our test set.}
\end{figure}

For the case of natural and bipartite networks, we find the LogNormal distribution to provide better fits than the Gamma or the Weibull distribution. We emphasize that this is an empirical finding which, to our knowledge, has not yet been justified theoretically. In this sense, the work presented in this paper can be seen as the first such justification because the LogNormal is obtained a limiting case of the generalized Gamma distribution which, as shown in appendix~\ref{sec:derivation}, provides a physically plausible model of distance distributions in networks. 

In fact, just as in the previous subsections, the generalized Gamma distribution is again found to provide the best overall fits to the data considered here. Qualitative examples of this behavior are shown in Fig.~\ref{fig:fits}.

We also evaluated how the different distributions perform in predicting the average shortest path length and compared empirical means to the means of fitted models. For the Weibull, the mean is given by $\lambda \Gamma(1+1/\kappa)$, the mean of the Gamma is $\eta \theta$, that of the LogNormal is $\exp(\mu -\xi^2/2)$, and for the generalized Gamma we have
\begin{equation}
\mathbb{E}\{t\} = \sigma \, \frac{\Gamma((\alpha+1) / \beta)}{\Gamma(\alpha/\beta)}.
\end{equation}

\begin{table}[t!]
\centering
\caption{\label{tab:msq} Mean squared errors for Fig.~\ref{fig:means}}
\begin{tabular}{lrrrr}
\toprule
 & $f_{WB}$ & $f_{GA}$ & $f_{LN}$ & $f_{GG}$ \\
 \midrule
bipartite & 1.665 & 1.382 & 0.731 & 0.414 \\
natural & 0.216 & 0.479 & 0.862 & 0.093 \\
social & 0.474 & 0.219 & 0.648 & 0.220 \\
\midrule
overall & 1.745 & 1.479 & 1.303 & 0.478 \\
\bottomrule
\end{tabular}
\end{table}

Figure~\ref{fig:means} shows scatter plots of predicted average shortest path lengths versus empirically determined ones for the networks in the KONECT collection. Visual inspection suggests that w.r.t.~predicting average path lengths, the Weibull performs worse than the Gamma which performs worse than the LogNormal which is outperformed by the generalized Gamma distribution. This is backed by Tab.~\ref{tab:msq} which lists mean squared errors for the data in the figure. Again, the generalized Gamma distribution performs best.

\subsection{Embedding Path Length Histograms in 3D}

\begin{figure}[t!]
\centering
\subfloat[synthetic networks]{\includegraphics[width=0.45\textwidth]{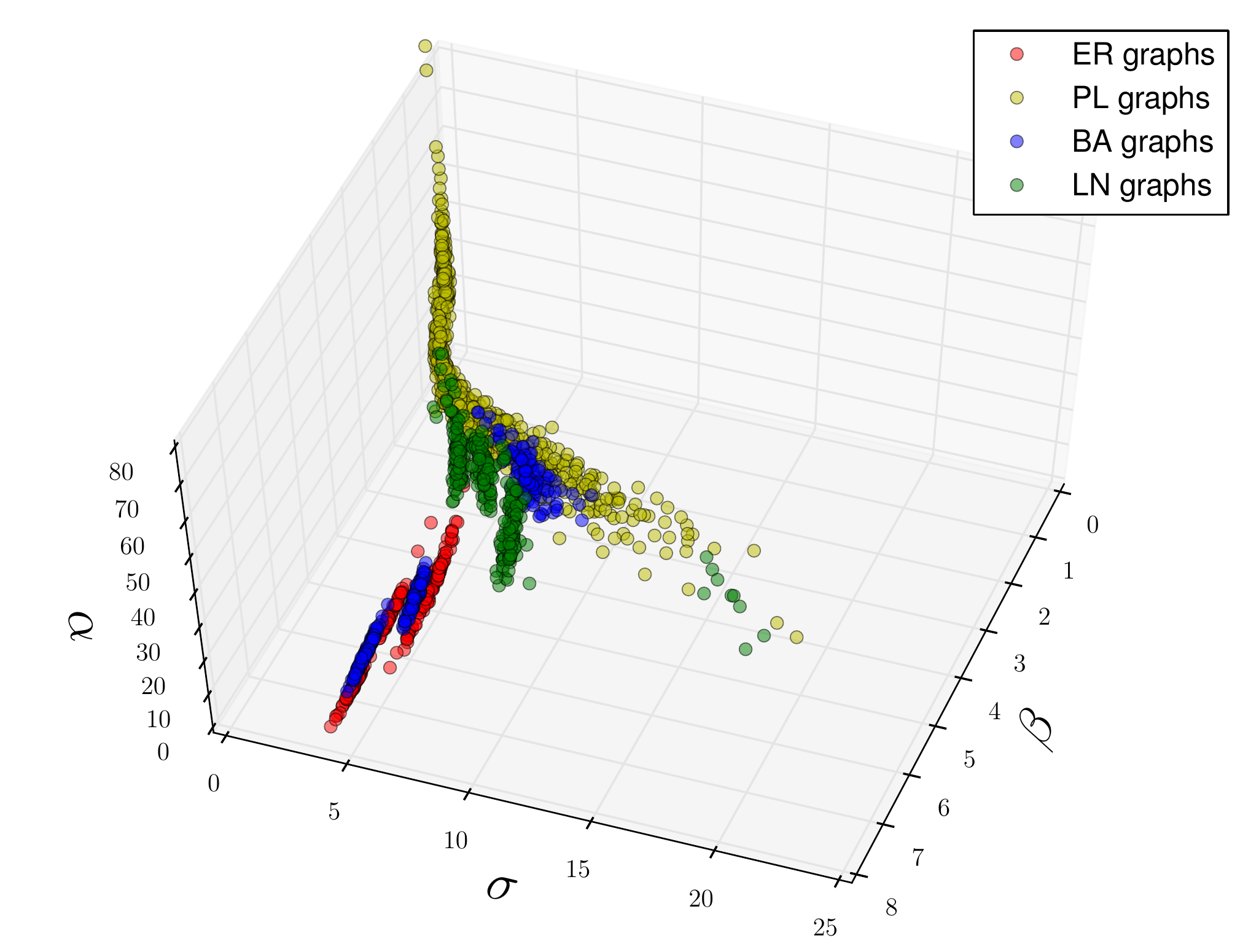}}

\subfloat[real-world networks]{\includegraphics[width=0.45\textwidth]{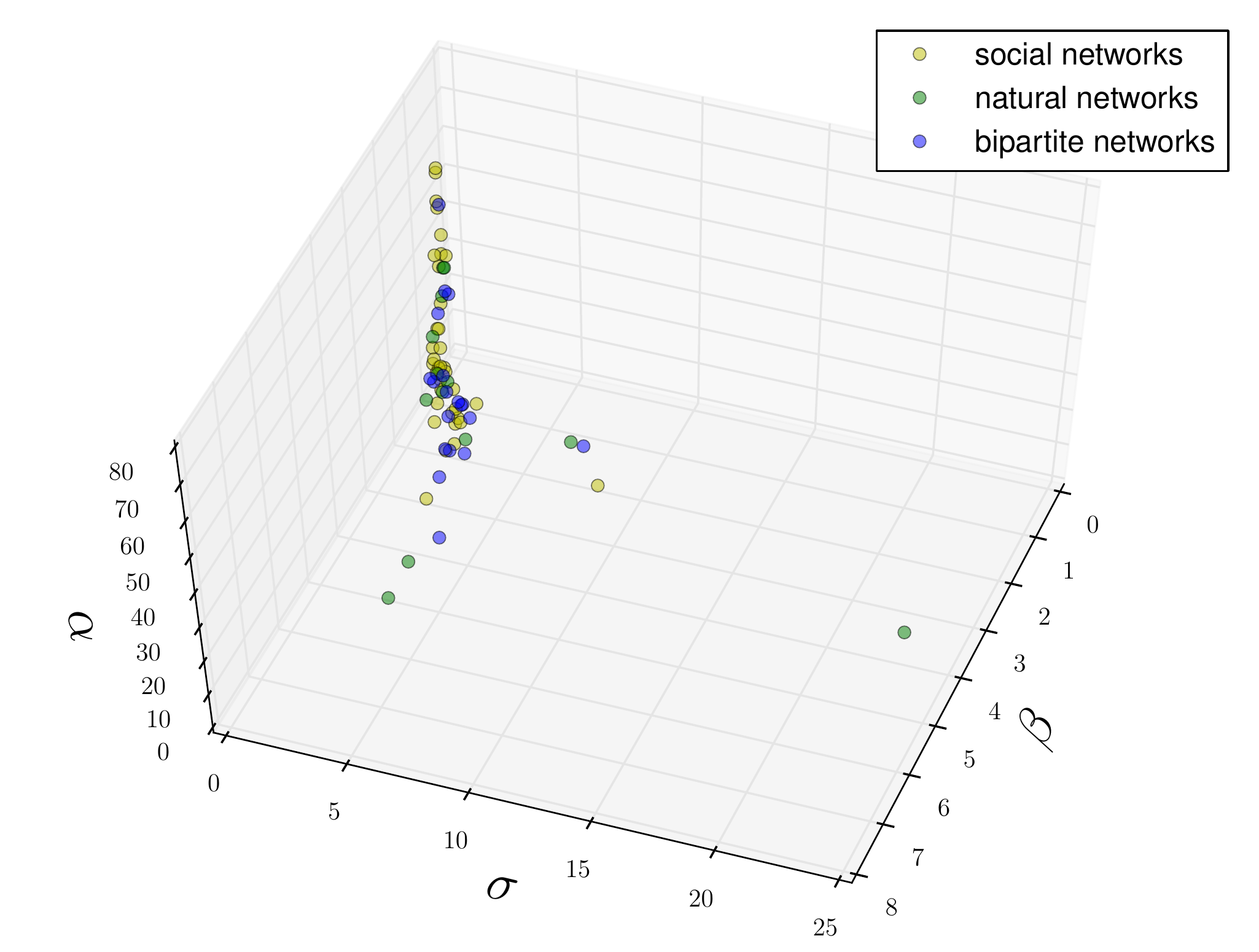}}
\caption{\label{fig:GGfits} 3D embedding of shortest path histograms from different networks. Each point $(\sigma, \alpha, \beta)$ represents a path length distribution in terms of the parameters of the best fitting generalized Gamma model. Path length distributions of different types of graphs appear to be confined to specific regions.}
\end{figure}

An interesting consequence of using the three-parameter generalized Gamma distribution to characterize shortest path histograms is that it provides a non-linear mapping of path length data into three dimensions. This allows for visual analytics of the behavior of different graph topologies w.r.t.~path length distributions.
Figure~\ref{fig:GGfits} shows exemplary shortest path distributions in terms 3D coordinates $(\sigma, \alpha, \beta)$ that result from fitting generalized Gamma distributions. Looking at the figure, it appears that shortest path distributions obtained from different network topologies cluster together or are confined to certain regions in this parameter space. These preliminary observations are arguably the most interesting finding in this paper as they suggest that the idea of characterizing networks in terms of continuous models of shortest path distributions can inform approaches to the problem of network inference from outbreak data. For now, we leave the study of the characteristics of these representations and their possible interpretations to future work.

\section{Summary and Outlook}
\label{sec:conclusion}

In this paper, we considered the problem of representing discrete path length histograms and epidemic outbreak data by means of continuous, analytically tractable models. 

Invoking the maximum entropy principle, we showed that generalized Gamma distribution provides a physically plausible model for these data, independent of the topology of the underlying networks. This result generalizes earlier models \cite{Bauckhage2013-TWA,Vazquez2006-PGI} and explains recent empirical observations made w.r.t.~twitter retweet networks \cite{Bild2014-ACO}. 

Empirical tests confirmed our theoretical prediction and revealed that the generalized Gamma distribution accounts well for shortest path histograms of synthesized Erd\H{o}s-R\'enyi, Barab\'asi-Albert, power law, and LogNormal graphs as well as for the empirical path length distributions of the real-world networks in the KONECT collection \cite{Kunegis2013-KON}.  

As an application in the context of visual analytics, we introduced a non-linear embedding of path length histograms into 3D parameter spaces and observed striking structural regularities in the resulting representations.

Given our results, there are several directions for future research. First of all, it appears worthwhile to attempt to relate the shape and scale parameters of the generalized Gamma distribution to physical properties or well established network features; we expect to be able to establish a connection to, say, the expected value or the variance of the generalized Gamma. 
Second of all, our theoretical results in this paper can be used to devised informed sampling schemes for the problem of computing shortest path histograms for large real-world networks. Efforts in this direction are underway and will hopefully be reported soon. 
Third of all, we are currently exploring ways of how to apply the results in Fig.~\ref{fig:GGfits} to the problem of network inference from outbreak data. A simple idea in this context is to create a large data base of 3D representations of outbreak data so that likely network structures for newly observed epidemic processes can be inferred from, say, nearest neighbor searches over the available examples.

\appendices

\section{Model Derivation}
\label{sec:derivation}

In this section, we show that the generalized Gamma distribution provides a principled model of shortest path- and outbreak statistics in networks. We argue based on Jaynes' \textit{maximum entropy principle} \cite{Jaynes1957-ITA} which states that, subject to observations and contextual knowledge, the probability distribution that best represents the available information is the one of highest entropy. In particular, we resort to an approach by Wallis (cf.~\cite[chapter 11]{Jaynes2003-PT}) which does not assume entropy as an a priori measure of uncertainty but uncovers it in the course of the argument.

To derive our analytical model of path length histograms of arbitrary networks, we consider the network spreading process in Fig.~\ref{fig:didactic-spread}. Borrowing terminology from epidemiology, we note that, at onset time $t=0$, most nodes in the network are \emph{susceptible} and one node is \emph{infected}. At time $t+1$, nodes that were infected at $t$ have \emph{recovered} yet did infect their susceptible neighbors. Highly infectious \emph{SIR} cascades like this realize a node discovery process: the number $n_t$ of newly infected nodes at time $t$ corresponds to the number of nodes that are at topological distance $d=t$ from the source. 

Given these considerations, we observe that the discrete shortest path distribution of the entire network is nothing but the sum of all node count histograms $h_s[t]$ taken over all possible source nodes.

\begin{figure}[t!]
\centering
\subfloat[constant $q_k$]{\includegraphics[width=0.3\columnwidth]{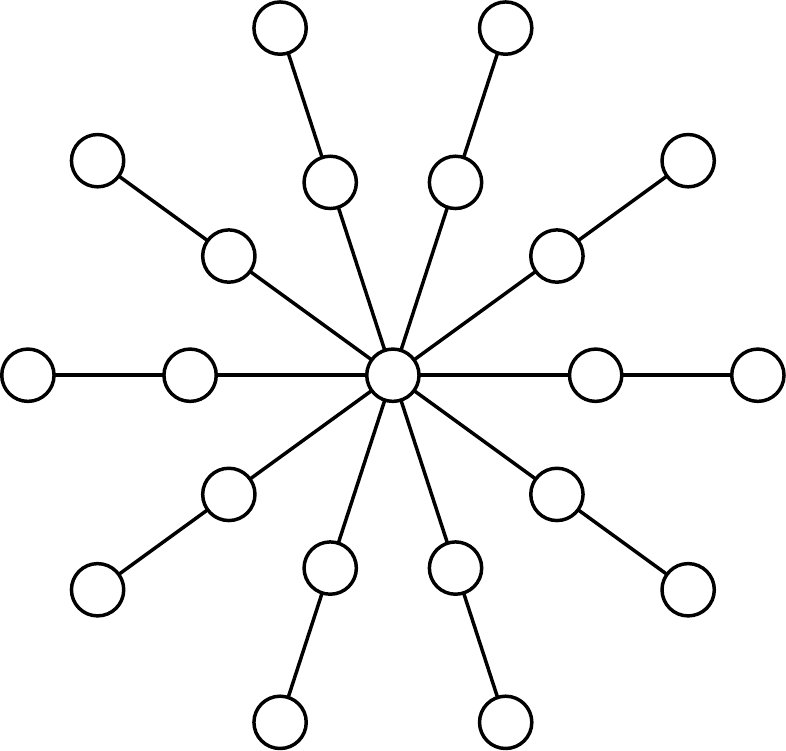}}\hfill
\subfloat[decreasing $q_k$]{\includegraphics[width=0.3\columnwidth]{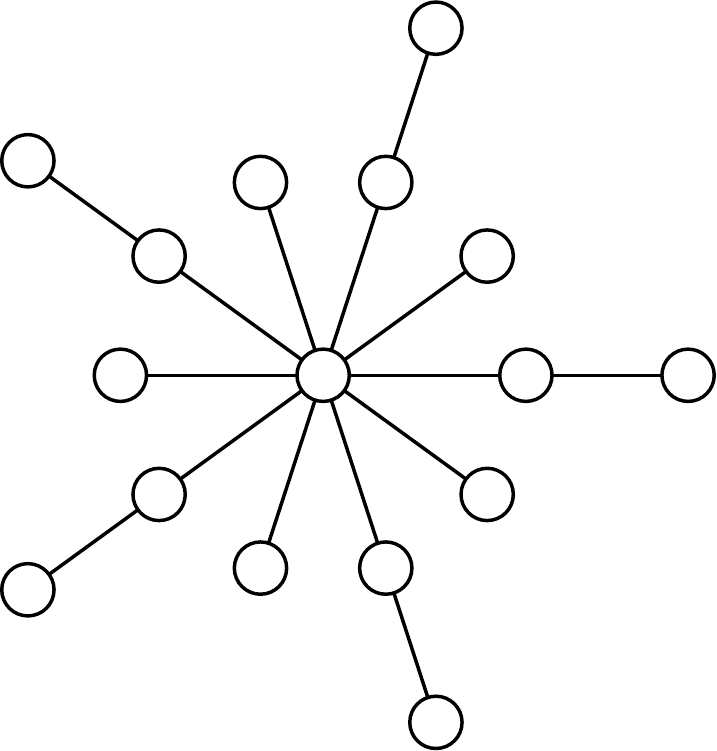}}\hfill
\subfloat[increasing $q_k$]{\includegraphics[width=0.3\columnwidth]{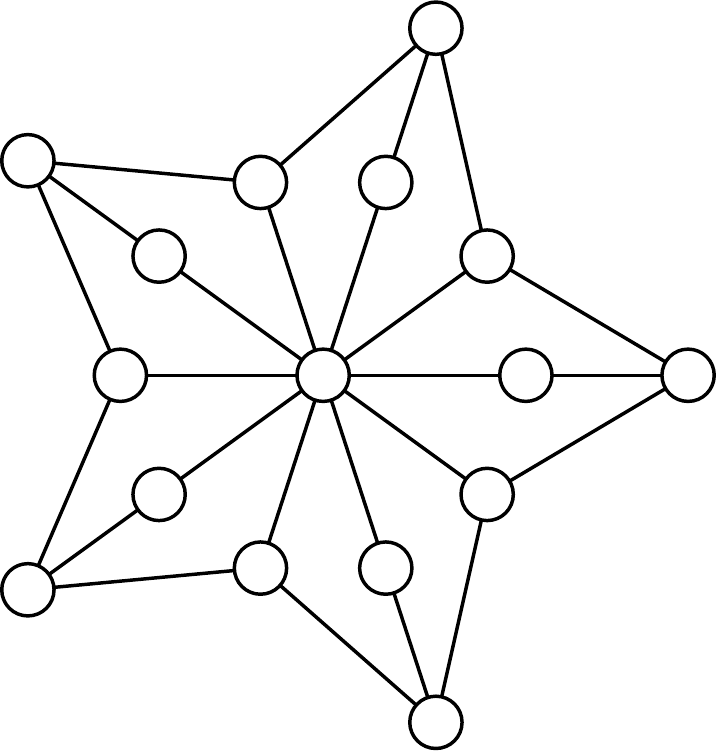}}
\caption{\label{fig:didactic} The probability of fining a path of length $k$ may be constant or decrease or increase with $k$.}
\end{figure}

In what follows, we let $K$ denote the length of the longest shortest path starting at a source $v_s$ and use $n_k$ to indicate the number of nodes that are $k$ steps away from $v_s$. Then,
\begin{equation}
n = \sum_{k=0}^K n_k
\end{equation}
where $n$ denotes the total number of nodes. The probability of observing a node at distance $k$ can thus be expressed as $p_k = n_k / n$ and we have
\begin{equation}
1 = \sum_{k=0}^K p_k.
\end{equation}
Moreover, we let $q_k$ denote the probability that there exists a shortest path of length $k$ so that
\begin{equation}
1 = \sum_{k=0}^K q_k
\end{equation}
and we emphasize that $p_k$ and $q_k$ will generally differ (see the didactic examples in Fig.~\ref{fig:didactic}).

With these definitions, the joint probability of observing counts $n_k$ of infected nodes corresponds to the multinomial 
\begin{equation}
P(n_1, \ldots, n_K) = n! \prod_{k = 1}^K \frac{q_k^{n_k}}{n_k!} 
\end{equation} 
whose log-likelihood is given by
\begin{equation}
\label{eq:logl1}
L = \log n! + \sum_{k=0}^K n_k \log q_k - \log n_k! 
\end{equation}
Assuming that $n_k \gg 1$, we may apply Stirling's formula $\log n_k! \approx n_k \, \log n_k - n_k$ to simplify \eqref{eq:logl1} which then becomes
\begin{align}
L & \approx n \, \log n - n + \sum_{k=0}^K n_k \bigl( \log q_k - \log n_k + 1 \bigr) \notag \\
  & = n \, \log n + n \sum_{k=0}^K p_k \bigl( \log q_k - (\log p_k + \log n) \bigr) \notag \\
  & = - n \sum_{k=0}^K p_k \, \log \frac{p_k}{q_k} \label{eq:MaxEnt}.
\end{align}    

As the expression in \eqref{eq:MaxEnt} is a Kullback-Leibler divergence,  
our considerations so far led indeed to an entropy that needs to be maximized in order to determine the most likely values $n_k^*$ of the $n_k$.

However, up until now, the quantities $q_k$ are not defined precisely enough to allow for a solution. Also, \eqref{eq:MaxEnt} accounts only indirectly for temporal aspects of network spreading processes as any dependency on time is hidden in the index of summation $k$. We address both these issues by choosing the ansatz 
\begin{equation}
\label{eq:p3}
q_k = A \, t_k^{\alpha-1}
\end{equation}
where $A$ is a normalization constant and $\alpha > 0$. This choice and its significance will be justified in detail in appendix~\ref{sec:alpha}. For now, we continue with our main argument.

Another underspecified quantity so far is the length $K$ of the supposed longest shortest path. However, dealing with networks of finite size, we can bypass the need of having to specify $K$ by means of introducing the following constraint 
\begin{equation}
\label{eq:p1}
\sum_{k=0}^{\infty} n_k = n 
\end{equation}
into the problem of maximizing \eqref{eq:MaxEnt}. Finally, to prevent degenerate solutions (e.g. instantaneous or infinite spread), we impose a constraint on the times $t_k$ and require their moments 
\begin{equation}
\label{eq:p2}
\sum_{k=0}^{\infty} \frac{n_k}{n} \, t_k^\beta = c 
\end{equation}
to be finite for some $\beta > 0$.

Using differential forms, we now express the log-likelihood in \eqref{eq:MaxEnt} and the two constraints in \eqref{eq:p1} and \eqref{eq:p2} as 
\begin{align*}
dL & = \sum_{k=0}^\infty \frac{\partial L}{\partial n_k} d n_k = \sum_{k=0}^\infty \left( \log A t_k^{\alpha - 1} - \log n_k \right) d n_k \\
dn & = \sum_{k=0}^\infty dn_k \\
dc & = \sum_{k=0}^\infty  t_k^\beta dn_k
\end{align*}
and consider the Lagrangian with multipliers $\rho$ and $\gamma$
\begin{equation}
\mathcal{L} = \sum_{k=0}^{\infty} \biggl [ \log \frac{A \, t_k^{\alpha-1}}{n_k} - \rho - \gamma t_k^\beta \biggr ] dn_k = 0
\end{equation}    
in order to determine most likely infection counts $n_k^*$ for the spreading process described above. Since at the solution the bracketed terms $[\cdot]$ must vanish identically for every $dn_k$, we immediately obtain
\begin{equation}
n_k^* = A \, e^{-\rho} \, t_k^{\alpha-1} \, e^{- \gamma t_k^\beta}.
\end{equation}
Plugging this result back into $p_k = n_k / n$ yields a discrete probability mass function 
\begin{equation}
\label{eq:preprefinal}
\frac{n_k^*}{n} = \frac{t_k^{\alpha-1} \, e^{- \gamma t_k^\beta}}{\sum\limits_{j=0}^\infty t_j^{\alpha-1} \, e^{- \gamma t_j^\beta}}
\end{equation}
which explicitly relates infection counts $n_k$ to time steps $t_k$. However, \eqref{eq:preprefinal} still depends on the Lagrangian multiplier $\gamma$ which is not immediately related to any available data.

Assuming the duration $\Delta t = t_{k+1} - t_k$ of time steps to be small, permits the approximation
\begin{equation*}
\sum_{k=0}^\infty t_k^{\alpha-1} \, e^{- \gamma t_k^\beta} \Delta t 
\approx \int_0^\infty t^{\alpha-1} \, e^{- \gamma t^\beta} dt
= \gamma^{-\frac{\alpha}{\beta}} \frac{\Gamma(\alpha/\beta)}{\beta}
\end{equation*}
so that
\begin{equation}
\frac{n_k^*}{n} = \Delta t \, \frac{\gamma^{\frac{\alpha}{\beta}} \beta}{\Gamma(\alpha/\beta)} \, t_k^{\alpha-1} \, e^{- \gamma t_k^\beta}.
\end{equation}

If this is plugged into \eqref{eq:p2}, tedious but straightforward algebra yields $\gamma = \tfrac{\alpha}{\beta c}$ which is to say that
\begin{equation}
\label{eq:prefinal}
\frac{n_k^*}{n} = \Delta t \biggl [ \frac{\beta}{\Gamma(\alpha/\beta)} \Bigl(\frac{\alpha}{\beta c}\Bigr)^\frac{\alpha}{\beta} \biggr ] t_k^{\alpha-1} e^{-\frac{\alpha}{\beta c} t_k^\beta}.
\end{equation}

In order to establish our final result, we now let $\Delta t \rightarrow 0$ which leads to 
\begin{equation}
\label{eq:limitproc}
\frac{n_k^*}{n} \approx \int_{\Delta t} f(\tau) \, d\tau \approx \Delta t \, f(t)
\end{equation}
where $t_k - \frac{\Delta t}{2} \leq t \leq t_k + \frac{\Delta t}{2}$. Direct comparison of \eqref{eq:prefinal} and \eqref{eq:limitproc} together with the substitution $\ggscl = \bigl(\tfrac{\beta c}{\alpha} \bigr)^{1/\beta}$ finally establishes that
\begin{equation}
f(t) = \frac{\beta}{\ggscl^\alpha} \frac{1}{\Gamma(\alpha/\beta)} t^{\alpha-1} e^{-(t/\ggscl)^\beta}
\end{equation}
which is indeed the generalized Gamma distribution that was introduced in equation \eqref{eq:GG}.

\section{Different Time Scales}
\label{sec:timescales}

Our derivation above was based on properties of networked $SIR$ cascades for which the infection rate $i$ as well as the recovery rate $r$ were both assumed to be 100\%. Yet, the empirical result in section~\ref{sec:experiments} indicate that the generalized Gamma distribution also account for the dynamics of less infectious spreading processes where $i$ and $r$ are smaller. 

Such epidemics usually last longer as it takes more time to reach all susceptible nodes and infected nodes may remain so over extended periods. In other words, such processes can be thought of as taking place on a different time scale. Here, we briefly show that the generalized Gamma distribution can also explain spreading processes on linearly or polynomially transformed time scales.

Recall that if a random variable $X$ is distributed according to $f(x)$, the monotonously transformed random variable $Y = h(X)$ has a probability function that is given by
\begin{equation}
g(y) = f \bigl( h^{-1}(y) \bigr) \cdot \left  \lvert \frac{d}{dy} h^{-1} (y) \right \rvert.
\end{equation}

Now, if $t$ is generalized Gamma distributed and $\tau = c t$ is a linearly transformed version of $t$, then 
\begin{equation}
t = \frac{\tau}{c} \quad \text{and} \quad \frac{dt}{d\tau} = \frac{1}{c}
\end{equation}
so that $\tau$ is distributed according to 
\begin{align*}
g(\tau) & = \frac{\beta}{\ggscl^\alpha} \frac{1}{\Gamma(\alpha/\beta)} \left(\frac{\tau}{c}\right)^{\alpha-1} e^{-(\tau / c \ggscl)^\beta} \cdot \frac{1}{c} \\
& = \frac{\beta}{\sigma'^\alpha} \frac{1}{\Gamma(\alpha/\beta)} \tau^{\alpha-1} e^{-(\tau/ \sigma')^\beta} 
\end{align*}
which is a generalized Gamma distribution with scale parameter $\sigma' = c \ggscl$.

By the same token, if $t$ is generalized Gamma distributed and $\tau = t^c$ is a polynomially transformed version of $t$, then 
\begin{equation}
t = \tau^\frac{1}{c} \quad \text{and} \quad \frac{dt}{d\tau} = \frac{1}{c} \, \tau^{\frac{1}{c} - 1}
\end{equation}
so that $\tau$ is distributed according to 
\begin{align*}
g(\tau) & = \frac{\beta}{\ggscl^\alpha} \frac{1}{\Gamma(\alpha/\beta)} \left(\tau^\frac{1}{c}\right)^{\alpha-1} e^{-(\tau^{1/c} / \ggscl)^\beta} \cdot  \frac{1}{c} \, \tau^{\frac{1}{c} - 1} \\
& = \frac{\beta' c}{\ggscl'^{\alpha'}} \frac{1}{\Gamma(\alpha'/\beta')} \tau^{\alpha'-1} e^{-(\tau^{\beta'} / \ggscl'^{\beta'})}
\end{align*}
which is a generalized Gamma distribution with  parameters $\ggscl' = \ggscl^c$, $\alpha' = \tfrac{\alpha}{c}$, and $\beta' = \tfrac{\beta}{c}$.

\section{Further Details}
\label{sec:alpha}

\begin{figure}[t!]
\centering
\includegraphics[width=0.9\columnwidth]{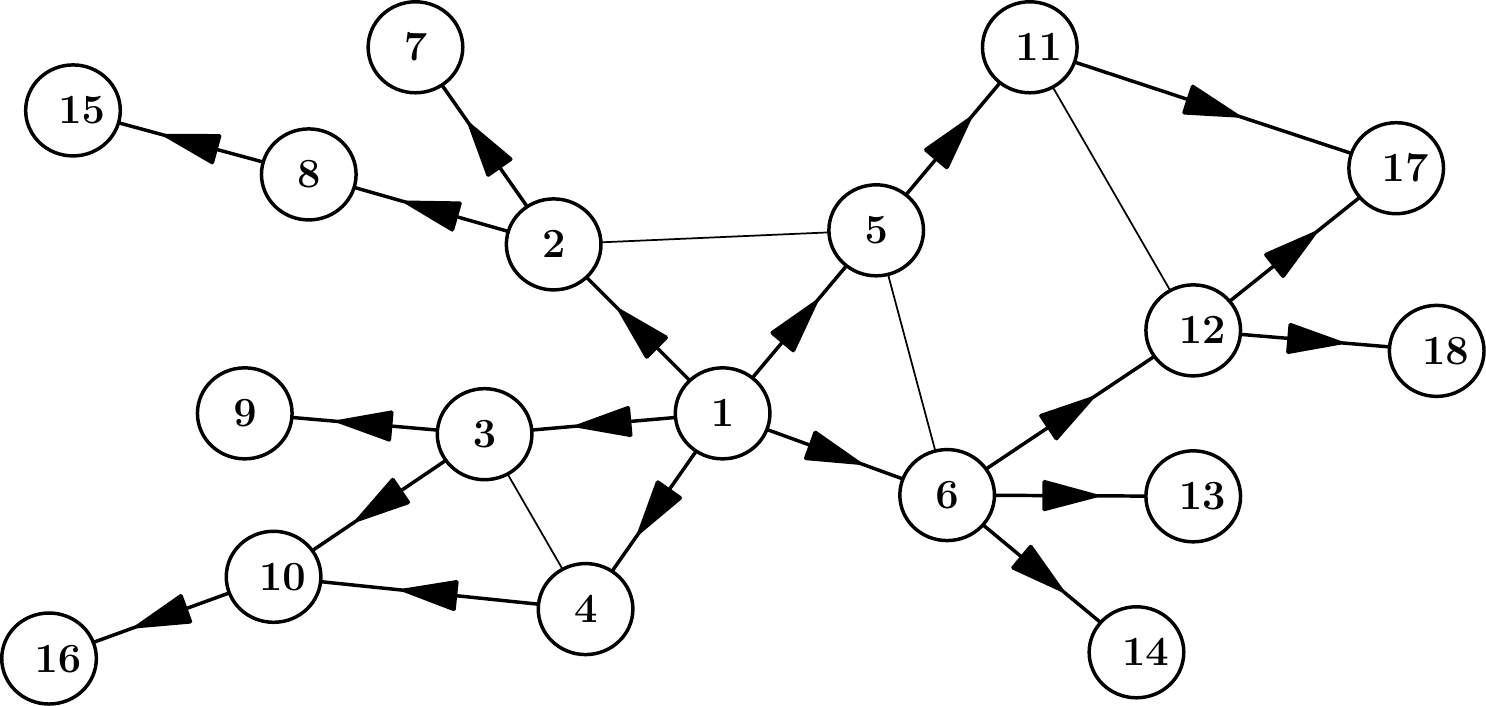}
\caption{\label{fig:didactic-trans} Causal graph of the process in Fig.~\ref{fig:didactic-spread}. The adjacency matrix of a directed acyclic graph with a single source node can be brought into strictly upper triangular form and is thus nilpotent.}
\end{figure}

While the constraints in (\ref{eq:p1}) and (\ref{eq:p2}) are arguably intuitive, the ansatz in (\ref{eq:p3}) merits further elaboration. 

Note that \emph{causal graphs} as in Fig.~\ref{fig:didactic-trans} indicate possible transmission pathways of an epidemic. In general they may be arbitrarily complex but, for the type of spreading process studied here, they are necessarily directed and acyclic.

Consider a network of $n$ nodes. If $v_s$ is the source node of an epidemic on this network and $\mat{A}$ is the adjacency matrix of the corresponding causal graph where
\begin{equation*}
A_{ij} = \begin{cases} 1 & \text{if nodes $v_i$ infects $v_j$} \\ 0 & \text{otherwise} \end{cases}
\end{equation*}
then $\bigl ( \mat{A}^k \bigr )_{sj}$ denotes number of ways the epidemic agent can reach $v_j$ from $v_s$ in $k$ steps and 
\begin{equation*}
q_k \propto \sum_{j} (\mat{A}^k)_{sj}.
\end{equation*}

For causal graphs, $\mat{A}$ is strictly upper triangular and therefore \textit{nilpotent}. That is, $\exists \, K \leq n : \mat{A}^k = \mat{0} \; \forall \,k > K$.

Hence, $\rho(\mat{A}) < 1$ which is to say $\exists \, u > 0 : \lVert \mat{A}^k \rVert_F < u \; \forall \, k$ and therefore
\begin{equation*}
0 \leq (\mat{A}^k)_{sj} \leq \lVert \mat{A}^k \rVert_F < u.
\end{equation*}
Accordingly, $\sum_{j} (\mat{A}^k)_{sj} \in O(k^{\alpha-1})$ for some $\alpha \geq 1$. Setting $t_k = \epsilon k$ where $\epsilon \leq 1$ yields $q_k \in O(t_k^{\alpha-1})$ which justifies \eqref{eq:p3}.

Put in simple terms, this is to say that it is sufficient to model the probability $q_k$ of observing an epidemic path of length $k$ as a polynomial in $t$ rather than, say, an exponential function.

The fact that the parameter $\alpha$ reflects aspects of network topology is easily seen from the examples in Fig.~\ref{fig:didactic}. Assuming the central node to be the source node of an epidemic the figure shows that $q_k$ may be constant ($\alpha = 1$), decreasing ($\alpha < 1$), or increasing ($\alpha > 1$) with $k$.

\bibliographystyle{IEEEtran}
\bibliography{literature}

\end{document}